\documentclass[aps,prd,onecolumn,preprintnumbers,notitlepage,superscriptaddress,
nofootinbib,amsfont,amssymb,10pt,singlespacing] {revtex4-1}
\usepackage{amsmath,bm}
\usepackage{times}
\usepackage{braket}
\usepackage{color,graphicx}
\usepackage{slashed}
\usepackage{hyperref}
\usepackage{graphics}
\usepackage{subfigure}
\usepackage{multirow,makecell}
\usepackage{textcomp}
\usepackage{mathtools}
\usepackage{float}
\newcommand{\beq}{\begin{eqnarray}}
\newcommand{\eeq}{\end{eqnarray}}

\def\lsim{ {\ \lower-1.2pt\vbox{\hbox{\rlap{$<$}\lower6pt\vbox{\hbox{$\sim$}
}}}\ } }
\def\gsim{ {\ \lower-1.2pt\vbox{\hbox{\rlap{$>$}\lower6pt\vbox{\hbox{$\sim$}
}}}\ } }
\definecolor{Red}{rgb}{1.,0.,0.}

\definecolor{Blue}{rgb}{0.,0.,1.}

\definecolor{nicered}{rgb}{0.7,0.1,0.1}
\definecolor{nicegreen}{rgb}{0.1,0.5,0.1}

\hypersetup{colorlinks,citecolor=nicegreen,linkcolor=nicered}

\begin{document}

\title{Quasi-two-body decays $B_c^+ \to \chi_{c0,c1} [\rho(K^*) \to] \pi\pi(K\pi)$ in the PQCD approach}

\author{Jun~Deng}
\email[Electronic address:]{1239039694@qq.com}
\affiliation{School of Physical Science and Technology,
 Southwest University, Chongqing 400715, China}

\author{Xian-Qiao~Yu}
\email[Electronic address:]{yuxq@swu.edu.cn}
\affiliation{School of Physical Science and Technology,
Southwest University, Chongqing 400715, China}

\date{\today}

\begin{abstract}
We study the quasi-two-body decays $B_c^+ \to \chi_{c0,c1} [\rho(K^*) \to]
\pi^+\pi^0(K^0\pi^+)$ in the perturbative QCD framework at leading order.
The $\pi\pi$ and $K\pi$ pairs are produced via the P-wave resonances
$\rho(770)$, $\rho(1450)$, $\rho(1700)$, and $K^*(892)$, parametrized
by the Gounaris-Sakurai and relativistic Breit-Wigner models.

The $\rho$ channels, dominated by the $\rho(770)$ resonance, yield
$\mathcal{B}(B_c^+ \to \chi_{c0} \pi^+\pi^0) = 7.33 \times 10^{-4}$
and $\mathcal{B}(B_c^+ \to \chi_{c1} \pi^+\pi^0) = 1.02 \times 10^{-3}$,
with constructive $\rho$ interference making the coherent total $\sim 26\%$--$30\%$ larger than the $\rho(770)$ contribution alone.
The Cabibbo-suppressed $K\pi$ channels yield branching ratios roughly a factor
of five below those of the $\pi\pi$ modes,
$\mathcal{B}(B_c^+ \to \chi_{c0} K^0\pi^+) = 1.49 \times 10^{-4}$
and $\mathcal{B}(B_c^+ \to \chi_{c1} K^0\pi^+) = 1.94 \times 10^{-4}$, the
Cabibbo suppression being only partially compensated by the narrow $K^*(892)$ resonance and
the normalization factor $N_{K^*}=1.48$. For $\chi_{c1}$
modes, $f_L \approx 93\%$ dominates and decreases with increasing resonance
mass. The ratio $R_{\chi_{c1}/\chi_{c0}}^{\pi\pi} \approx 1.38$ reflects the
common resonant $\rho$ production mechanism shared by $\chi_{c0}$ and
$\chi_{c1}$, with the charmonium decay-constant hierarchy and phase space
determining the relative yield of P-wave charmonia. The analogous ratio in the Cabibbo-suppressed $K\pi$ channel, $R_{\chi_{c1}/\chi_{c0}}^{K\pi} \approx 1.30$, is consistent with this value, confirming that the spin ratio is governed by charmonium-side dynamics.
\end{abstract}
\maketitle

%
%
\section{Introduction}\label{sec:intro}
The \(B_c\) meson, composed of two different heavy quarks (\(b\) and \(\bar{c}\)), occupies a unique position in the Standard Model of particle physics. As the lightest \(\bar{b}c\) bound state, it was first discovered by the CDF Collaboration at the Tevatron collider~\cite{CDF:1998,Chang:1994}. The LHCb experiment has since measured several key ratios, including \(R_{K/\pi} \equiv \mathcal{B}(B_c^+ \to J/\psi K^+)/\mathcal{B}(B_c^+ \to J/\psi \pi^+) = 0.079 \pm 0.007 \pm 0.003\)~\cite{LHCb:2016} and \(R_{2\pi} \equiv \mathcal{B}(B_c^+ \to J/\psi \pi^+\pi^0)/\mathcal{B}(B_c^+ \to J/\psi \pi^+) = 2.80 \pm 0.25\)~\cite{LHCb:2023}. Unlike heavy-light mesons such as \(B\) or \(D\) mesons, the \(B_c\) meson cannot decay via strong interactions and is stable against strong decays, resulting in a relatively long lifetime. The distinctive feature involving two heavy quarks with different flavors allows either of the heavy quarks to decay with the other behaving as a spectator quark, or both quarks to annihilate into a virtual \(W^+\) boson. Therefore, its rich decay channels, involving the weak decays of either the \(b\) or \(c\) quark as well as annihilation channels, make it an ideal laboratory for testing the Standard Model and probing non-perturbative Quantum Chromodynamics (QCD) dynamics, which might shed light on possible new physics beyond the Standard Model (SM)~\cite{Li:2018}.

In particular, the exclusive decays of the \(B_c\) meson into P-wave charmonia (\(\chi_{cJ}\) with \(J = 0, 1, 2\)) provide a unique probe into heavy quarkonium production mechanisms, as the formation of P-wave states involves non-trivial orbital angular momentum contributions. Theoretical investigations into \(B_c \to \chi_{cJ}\) transitions~\cite{Liu:2018} have revealed significant challenges within the conventional collinear factorization frameworks, especially after the recent LHCb observation of \(B_c^+ \to \chi_{c2}\pi^+\), the upper limit on \(B_c^+ \to \chi_{c1}\pi^+\)~\cite{LHCb:2024}, and evidence for \(B_c^+ \to \chi_{c0}\pi^+\)~\cite{LHCb:2016chi}. A prominent issue arises in the decay to the axial-vector charmonium \(\chi_{c1}\) (\(J^{PC}=1^{++}\)): its leading-twist decay constant vanishes in the limit of heavy quark symmetry, causing the naive factorization approach to fail in accurately describing processes involving \(\chi_{c1}\)~\cite{Liu:2018}. To resolve this, one must go beyond the leading-twist approximation by including higher-twist light-cone distribution amplitudes (LCDAs). The perturbative QCD (PQCD) factorization approach~\cite{Keum:2001,Lu:2001,Li:2003,Xiao:2014}, which retains the transverse momentum \(k_T\) of the valence quarks, offers a consistent framework to handle these difficulties by effectively regulating the endpoint singularities and providing non-vanishing contributions for \(\chi_{c1}\) production. The case of \(\chi_{c0}\) (\(J^{PC}=0^{++}\)) presents a different challenge: its production via a color-singlet current is suppressed in the naive factorization approach, and understanding the non-factorizable contributions to its production has become a central topic in heavy quarkonium decays~\cite{Beneke:1999,Beneke:2000}.

Recent improved perturbative QCD (iPQCD) calculations of the two-body decays \(B_c^+ \to \chi_{cJ}\pi^+\)~\cite{Liu:2023,Liu:2025} have revealed a significant tension with experiment for the \(\chi_{c0}\) channel: while the predicted ratio \(R_{\chi_{c0}/J/\psi} = 0.17 \pm 0.02\) is substantially lower than the value \(1.41^{+0.50}_{-0.45}\) inferred from LHCb measurements~\cite{LHCb:2016chi}. A peculiar interference pattern between twist-2 and twist-3 contributions has been identified in the \(\chi_{c1}\) modes, but not in the corresponding \(\chi_{c0,c2}\) channels~\cite{Liu:2025}. This motivates a more detailed investigation of \(\chi_{c0}\) and \(\chi_{c1}\) production mechanisms, particularly in multi-body decay channels where resonant contributions may play a different role.

For the theoretical inputs required in our analysis, significant progress has been made in the calculation of \(B_c \to \chi_{cJ}\) transition form factors in various theoretical frameworks~\cite{Liu:2025}. Model-independent relations among the form factors for \(B_c \to P\)-wave charmonia have been derived using heavy quark spin symmetry (HQSS)~\cite{Liu:2018}, providing additional constraints on theoretical uncertainties. Complementary calculations based on QCD sum rules~\cite{Zhang:2023} and on nonrelativistic QCD (NRQCD)~\cite{Zhu:2018} have also been performed to systematically compute the form factors for \(B_c \to \chi_{cJ}\) transitions.

It is known that a factorization formalism describing multi-body \(B\) meson decays in the entire phase space is not yet available. However, when two final-state mesons are collimated and the bachelor meson recoils back, the quasi-two-body approximation provides a reasonable working framework~\cite{Chen:2003}. This situation occurs particularly in the low invariant mass region of the Dalitz plot~\cite{Dalitz:1953,Dalitz:1954}, where most resonant structures are seen. Within this framework, the non-perturbative dynamics responsible for the production of the meson pair, including final-state interactions, are absorbed into two-meson distribution amplitudes (DAs)~\cite{Muller:1994,Diehl:1998,Diehl:2000,Pire:2003}, which have been developed within the PQCD framework. The formulation of three-body decays is then simplified to that of quasi-two-body decays, where the hard kernel contains a single virtual gluon exchange at leading order in \(\alpha_s\). The typical PQCD factorization formula for the \(B_c \to P_1 P_2 P_3\) decay amplitude can be written as

\begin{equation}
	\mathcal{A}(B_c \to P_1 P_2 P_3) = \Phi_{B_c} \otimes H \otimes \Phi_{P_1 P_2} \otimes \Phi_{P_3},
\end{equation}

where the hard kernel \(H\) describes the dynamics of the strong and electroweak interactions, containing one hard gluon at leading order. The \(\Phi_{B_c}\) and \(\Phi_{P_3}\) are the wave functions for the \(B_c\) meson and the bachelor final-state meson, respectively, which absorb the non-perturbative dynamics in the relevant processes. The two-meson DA \(\Phi_{P_1 P_2}\) absorbs the non-perturbative dynamics associated with the production of the meson pair. In recent years, several theoretical approaches have been developed for describing three-body hadronic \(B\) meson decays, including analyses based on flavor symmetry principles~\cite{Gronau:2013,Gronau:2005,Engelhard:2005,Imbeault:2011,Xu:2014,He:2015,Cheng:2007}, the QCD factorization (QCDF) approach~\cite{Furman:2005,El-Bennich:2006,Dedonder:2011,Cheng:2013,Cheng:2016,Klein:2017,Zhang:2013,Qi:2019}, and the perturbative QCD (PQCD) approach~\cite{Ali:2007,Li:2017a,Li:2017,Ma:2017,Wang:2016,Rui:2018,Zou:2020,Li:2021,Yan:2022}. The quasi-two-body PQCD framework has been successfully applied to various \(B\) and \(B_c\) three-body decays~\cite{Chen:2003,Ali:2007,Li:2017,Wang:2026}. Moreover, four-body \(B\) meson decays have also been investigated in the quasi-two-body framework~\cite{Yan:2023,Yan:2024,Yan:2025,Liang:2022}. The two-meson P-wave DAs are parametrized and normalized to time-like form factors, which take the RBW model for the narrow \(K^*\) resonance and the GS model for the broad \(\rho\) resonance series. Following Ref.~\cite{Li:2017}, normalization factors $N_\rho=1.05$ and $N_{K^*}=1.48$ are introduced to remedy possible theoretical mismatches between the form factors and the resonance parameters. Such a parametrization is justified by the Watson theorem~\cite{Watson:1952}.

The quasi-two-body decays $B_c \to J/\psi \pi\pi(K\pi)$ via $\rho(K^*)$
resonances have recently been studied in the PQCD approach~\cite{Wang:2026},
revealing large longitudinal polarization fractions and constructive $\rho$
interference. The two-body decays $B_c \to J/\psi$ plus axial-vector or tensor
mesons have also been analyzed within the same framework~\cite{Zhao:2025}.

While the decay topology in the present work---factorizable and nonfactorizable
emission through an intermediate $\rho/K^*$ resonance---is structurally
identical to that of $B_c \to J/\psi \pi\pi(K\pi)$~\cite{Wang:2026}, the
replacement of the S-wave charmonium $J/\psi$ by P-wave charmonia $\chi_{c0,c1}$
introduces several new features. First, $\chi_{c0}$ ($J^{PC}=0^{++}$)
production via a color-singlet current is formally suppressed in naive
factorization, whereas $J/\psi$ production is not~\cite{Beneke:1999,Beneke:2000}.
Second, the axial-vector $\chi_{c1}$ ($J^{PC}=1^{++}$) has a vanishing
leading-twist decay constant in the heavy quark limit~\cite{Liu:2018}, so its
production must proceed through higher-twist LCDAs; no such suppression exists
for $J/\psi$. Third, the $\chi_{c1}$ channels carry three independent
polarization states ($L$, $\parallel$, $\perp$) whose interference pattern
is sensitive to the twist-2/twist-3 interplay, an
aspect absent for the purely vector $J/\psi$ final state. These differences
warrant a dedicated analysis rather than a direct
extrapolation of the $J/\psi$ results.

In this work, we extend the PQCD factorization approach to the quasi-two-body decays $B_c^+ \to \chi_{c0,c1} [\rho(K^*) \to] \pi^+\pi^0(K^0\pi^+)$. The $\pi\pi$ and $K\pi$ systems are described by the two-meson DAs in the P-wave, with the time-like form factors parametrized via the GS model for the broad $\rho$ resonances and the RBW model for the narrow $K^*$. The normalization factors $N_{\rho}$ and $N_{K^*}$ are taken from Ref.~\cite{Li:2017}. For the $\chi_{cJ}$ mesons, both twist-2 and twist-3 LCDAs are included. We compute the CP-averaged branching ratios for all channels, the polarization fractions $f_L,f_\parallel,f_\perp$ and individual polarization components for the $\chi_{c1}$ modes. Several relative ratios, including the spin ratio $R_{\chi_{c1}/\chi_{c0}}^{\pi\pi}$ and its Cabibbo-suppressed counterpart $R_{\chi_{c1}/\chi_{c0}}^{K\pi}$, the SU(3)-breaking ratio $R_{K/\pi}$, and $R_{J/\psi}^{\chi_{c0,c1}}$, are investigated and compared with naive factorization expectations and existing LHCb data~\cite{LHCb:2023}. The constructive $\rho$ interference pattern and its impact on the total branching ratios are also examined. Under the narrow-width approximation, two-body branching fractions for $B_c^+ \to \chi_{cJ} \rho^+(K^{*+})$ are extracted, enabling a direct comparison with the two-body iPQCD calculation of Ref.~\cite{Liu:2023,Liu:2025}.

The rest of this paper is organized as follows. In Sec.~\ref{sec:pert}, we define the kinematic variables and present the two-meson P-wave DAs for the \(\pi\pi\) and \(K\pi\) systems, including the parametrization of the time-like form factors. The PQCD decay amplitudes and numerical input parameters are given in Sec.~\ref{sec:numer}, followed by our results for branching ratios, polarization fractions, and relative ratios. Finally, a summary is presented in Sec.~\ref{sec:conclusion}.

%
%
\section{Theoretical Framework}\label{sec:pert}
\subsection{Kinematics}
In the quasi-two-body decays $B_c^+ \to \chi_{cJ} [\rho(K^*) \to] \pi\pi(K\pi)$ considered in this work, the underlying quark-level transition is the Cabibbo-favored $b \to c$ decay, specifically $\bar{b} \to \bar{c} u \bar{d}$ for the $\pi\pi$ channels and $\bar{b} \to \bar{c} u \bar{s}$ for the $K\pi$ channels. The corresponding low-energy effective Hamiltonian~\cite{Buchalla:1996} is

\begin{equation}
{\cal H}_{\rm eff}=\frac{G_{F}}{\sqrt{2}} V_{cb}^{*} V_{uq}\big[C_{1}(\mu)O_{1}^{(c)}(\mu)+C_{2}(\mu)O_{2}^{(c)}(\mu)\big],
\label{eq:heff}
\end{equation}

where $q = d, s$, $G_{F}=1.1663788\times10^{-5}~{\rm GeV^{-2}}$ is the Fermi constant, and $C_{1,2}$ are the Wilson coefficients. The tree-level current-current operators for the $b \to c$ transition are

\begin{equation}
\begin{split}
O_{1}^{(c)}=(\overline{b}_{\alpha}c_{\beta})_{V-A}(\overline{u}_{\beta}q_{\alpha})_{V-A},\qquad
O_{2}^{(c)}=(\overline{b}_{\alpha}c_{\alpha})_{V-A}(\overline{u}_{\beta}q_{\beta})_{V-A},
\end{split}
\end{equation}

where the subscripts $\alpha$ and $\beta$ are the color indices. The left-handed current is defined as $(\overline{b}_{\alpha}c_{\beta})_{V-A}=\overline{b}_{\alpha}\gamma_{\mu}(1-\gamma_{5})c_{\beta}$.

Penguin operators ($O_3$--$O_{10}$), which mediate flavor-changing neutral-current transitions $b\to d$ and $b\to s$, do not contribute to the charged-current $b\to c$ transition relevant for charmonium production in the decays considered here, and are therefore absent from Eq.~\eqref{eq:heff}. Consequently, only the tree operators $O_{1,2}^{(c)}$ are retained in our numerical analysis. Because these decays are then driven by a single tree-level amplitude carrying a real CKM factor $V_{cb}^*V_{uq}$, direct $CP$ violation is naturally absent in the considered quasi-two-body decays.

\begin{figure}
	\centering
	 \includegraphics{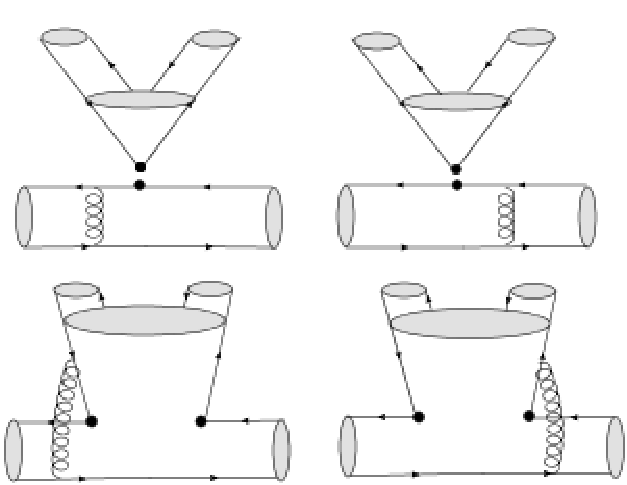}
	\caption{Feynman diagrams for the quasi-two-body decays $B_c^+ \to \chi_{cJ} [V\to] P_1P_2$ with $J=0,1$, $V=\rho,K^*$ and $P_1P_2=\pi\pi,K\pi$. Only the $J=0,1$ cases are considered in this work. The upper two diagrams are factorizable emission diagrams, and the lower two are nonfactorizable emission diagrams.}
	\label{fig:fig1}
\end{figure}

The relevant CKM factors in the decay amplitudes are $V_{cb}^* V_{ud}$ for the $\pi\pi$ channels and $V_{cb}^* V_{us}$ for the $K\pi$ channels, where $|V_{cb}| = 4.182 \times 10^{-2}$, $|V_{ud}| = 0.97435$, and $|V_{us}| = 0.22500$ are taken from the PDG 2024~\cite{PDG:2024}. Annihilation diagrams, where the $B_c^+$ annihilates into a virtual $W^+$ boson, are not included in the present analysis. These contributions are expected to be power-suppressed in the heavy quark limit, but their quantitative impact in $B_c$ decays may be non-negligible~\cite{Ivanov:2006,Ebert:2010} and warrants a dedicated study beyond the scope of this work. The relevant Feynman diagrams for the decay amplitudes are illustrated in Fig.~\ref{fig:fig1}.
In this paper, we work in the light-cone coordinate system. Taking the $B_c$ meson at rest as the center-of-mass frame, the light-cone coordinates are defined as follows:
	\begin{equation}
\begin{split}
p^{+}=\dfrac{p^{0}+p^{3}}{\sqrt{2}},p^{-}=\dfrac{p^{0}-p^{3}}{\sqrt{2}},p_{T}=(p^{1},p^{2}),
\end{split}
\end{equation}
where $p^{2}$, $p_{1}\cdot p_{2}$ are denoted as
	\begin{equation}
\begin{split}
p^{2}=2p^{+}p^{-}-p_{T}^{2},p_{1}\cdot p_{2}=p^{+}_{1}p^{-}_{2}+p^{-}_{1}p^{+}_{2}-p_{1T}\cdot p_{2T},
\end{split}
\end{equation}

In the quasi-two-body decay process $B_{c}^{+}\rightarrow \Psi(V\rightarrow)P_{1}P_{2}$ considered in this paper, we define the intermediate vector resonance $V$ with momentum $p$, and the three final-state mesons $\Psi$, $P_{1}$, $P_{2}$ with momenta $q$, $p_{1}$, $p_{2}$, respectively. The mass of the $B_{c}^{+}$ meson is denoted by $M_{B}$, and momentum conservation implies $p_{B}=p+q$ and $p=p_{1}+p_{2}$. We choose the meson $\Psi$ and the final-state $V$ to move along the directions $n=(1,0,0_{T})$ and $v=(0,1,0_{T})$, respectively. We also need to define the three valence quark momenta $k_{B}$, $k_{p}$, $k_{q}$ to evaluate the hard kernels $H$ in the PQCD approach.
\begin{equation}
\begin{split}
&p_{B}=\frac{m_{B_{c}}}{\sqrt{2}}(1,1,0_{T}),  k_{B}=(0,\frac{m_{B_{c}}}{\sqrt{2}}x_{B},k_{BT}),          \\
&p=\frac{m_{B_{c}}}{\sqrt{2}}(f^{-},f^{+},0_{T}),k_{q}=\frac{m_{B_{c}}}{\sqrt{2}}(zf^{-},zf^{+},k_{2T}),       \\
&q=\frac{m_{B_{c}}}{\sqrt{2}}(g^{+},g^{-},0_{T}),k_{p}=\frac{m_{B_{c}}}{\sqrt{2}}(x_{3}g^{+},x_{3}g^{-},k_{1T}),              \\
\end{split}
\end{equation}
The variables $g^{\pm}$ and $f^{\pm}$ read
\begin{equation}
\begin{split}
& f^{\pm}= \frac{1}{2}(1+\eta-r_{3}\pm \sqrt{(1-\eta)^{2}-2r_{3}(1+\eta)+r_{3}^{2}}), \\
& g^{\pm}= \frac{1}{2}(1-\eta+r_{3}\pm \sqrt{(1-\eta)^{2}-2r_{3}(1+\eta)+r_{3}^{2}}), \\
\end{split}
\end{equation}

We define $\eta=\omega^{2}/m_{B_c}^{2}$ and $r_{3}=m_{\Psi}^{2}/m_{B_c}^{2}$, where $\omega^{2}=p^{2}$ is the invariant mass squared of the meson pair and $m_{\Psi}$ is the charmonium mass. We denote by $x_{B}$, $z$, and $x_{3}$ the momentum fractions of the light quark in each meson, each ranging from $0$ to $1$. The related polarization vectors of the P-wave pairs and $\chi_{c1}$ meson can then be defined as: 
\begin{equation}
	\begin{split}
		\epsilon_{p}^{L}=\frac{1}{\sqrt{2\eta}}(-f^{-},f^{+},0_{T}),\epsilon_{p}^{T}=(0,0,1_{T}), \\    \epsilon_{q}^{L}=\frac{1}{\sqrt{2r_{3}}}(g^{+},-g^{-},0_{T}),\epsilon_{q}^{T}=(0,0,1_{T}), \\           
	\end{split}
\end{equation}
satisfying the normalization $\epsilon_{p}^{2}=\epsilon_{q}^{2}=-1$ and the orthogonality $\epsilon_{p}\cdot p=\epsilon_{q}\cdot q=0$.

According to the relation $p=p_{1}+p_{2}$ and the on-shell conditions $p^{2}_{1,2}=m_{P_{1,2}}^{2}$, $m_{P_{1,2}}$ being the $P_{1,2}$ meson masses, we can define momenta of $P_{1,2}$,

\begin{equation}
\begin{split}
&
p_{1}=((1-\zeta+\frac{r_{1}-r_{2}}{2\eta})\frac{m_{B_{c}}}{\sqrt{2}}f^{-},(\zeta+\frac{r_{1}-r_{2}}{2\eta})\frac{m_{B_{c}}}{\sqrt{2}}f^{+},P_{T}),\\
& p_{2}=((\zeta-\frac{r_{1}-r_{2}}{2\eta})\frac{m_{B_{c}}}{\sqrt{2}}f^{-},(1-\zeta-\frac{r_{1}-r_{2}}{2\eta})\frac{m_{B_{c}}}{\sqrt{2}}f^{+},-P_{T}), \\
& 
p_{T}^{2}=\zeta(1-\zeta)\omega^{2}+\frac{(m_{P_{1}}^{2}-m_{P_{2}}^{2})^{2}}{4\omega^{2}}-\frac{m_{P_{1}}^{2}+m_{P_{2}}^{2}}{2},\\
\end{split}
\end{equation}
 $r_{1,2}=m_{P_{1,2}}^{2}/m_{B_{c}}^{2}$ are the mass-squared ratios, and $\zeta+\frac{r_{1}-r_{2}}{2\eta}=p_{1}^{+}/p^{+}$ is the momentum fraction of one meson of the pair, up to corrections from the final-state masses. Alternatively, one may introduce the polar angle $\theta$ of the meson $P_{1}$ in the rest frame of the $P_{1}P_{2}$ pair. The relationship between the momentum fraction $\zeta$ and the polar angle $\theta$ in the dimeson rest frame can be derived straightforwardly:
\begin{equation}
	\begin{split}
	        2\zeta-1=\sqrt{1-2\frac{r_{1}+r_{2}}{\eta}+\frac{(r_{1}-r_{2})^{2}}{\eta^{2}}}\cos \theta,    \\
	\end{split}
\end{equation}
with the limits
\begin{equation}
	\begin{split}
		\zeta_{max,min}=\frac{1}{2}[1\pm\sqrt{1-2\frac{r_{1}+r_{2}}{\eta}+\frac{(r_{1}-r_{2})^{2}}{\eta^{2}}}]. \\
	\end{split}
\end{equation}
The relevant branching fraction for the quasi-two-body $B_{c}$ meson decay under consideration is expressed as~\cite{Li:2003}
\begin{equation}
	\begin{split}
	 \mathcal{B} = \frac{\tau_{B_c} m_{B_c}}{256\pi^3} \int_{(\sqrt{r_1}+\sqrt{r_2})^2}^{1} d\eta\, \sqrt{(1-\eta)^2 - 2r_3(1+\eta) + r_3^2} \int_{\zeta_{\min}}^{\zeta_{\max}} d\zeta\, |\mathcal{A}|^2,
	\end{split}
\end{equation}
where $\tau_{B_c}$ denotes the lifetime of the $B_{c}$ meson.

Angular momentum conservation forces the vector mesons $V$ in the quasi-two-body decays $B_{c}\rightarrow \chi_{c0}[V\rightarrow]P_{1}P_{2}$ to be fully longitudinally polarized. For the decays $B_{c}\rightarrow \chi_{c1}[V\rightarrow]P_{1}P_{2}$, both longitudinal and transverse polarizations contribute. The amplitudes can be decomposed as
\begin{equation}
	\begin{split}
	\mathcal{A} = \mathcal{A}_L + \mathcal{A}_N \epsilon_T \cdot \epsilon_{3T} + i \mathcal{A}_T \epsilon_{\alpha\beta\rho\sigma} n_+^\alpha n_-^\beta \epsilon_T^\rho \epsilon_{3T}^\sigma,
	\end{split}
\end{equation}
where $\mathcal{A}_L$ represents the longitudinally polarized decay amplitude, while $\mathcal{A}_N$ and $\mathcal{A}_T$ correspond to the transverse contributions. Consequently, the total decay amplitude for $B_{c}\rightarrow \chi_{c1}[V\rightarrow]P_{1}P_{2}$ decays can be written as
\begin{equation}
	\begin{split}
		|\mathcal{A}|^2 = |\mathcal{A}_0|^2 + |\mathcal{A}_\parallel|^2 + |\mathcal{A}_\perp|^2,
	\end{split}
\end{equation}
with $\mathcal{A}_0$,$\mathcal{A}_\parallel$ and $\mathcal{A}_\perp$ defined by
\begin{equation}
	\begin{split}
	\mathcal{A}_0 = \mathcal{A}_L, \quad \mathcal{A}_\parallel = \sqrt{2}\mathcal{A}_N, \quad \mathcal{A}_\perp = \sqrt{2}\mathcal{A}_T.
	\end{split}
\end{equation}
The corresponding polarization fractions $f_h$ are then given by
\begin{equation}
	\begin{split}
	f_h = \frac{\mathcal{B}_h}{\mathcal{B}_0 + \mathcal{B}_\parallel + \mathcal{B}_\perp}, \quad h = 0, \parallel, \perp
	\end{split}
\end{equation}
satisfying the normalization condition $f_0+f_\parallel+f_\perp=1$.

\subsection{Distribution Amplitudes}
For the $B_{c}$ meson wave function, we concentrate on the leading-power contribution and adopt the form commonly used in perturbative QCD calculations~\cite{Xiao:2012,Li:2003,Liu:2018,Liu:2023}:
\begin{equation}
	\begin{split}
		\Phi_{B_c} = \frac{i}{\sqrt{2N_c}}(\not\! p_{B_c} + m_{B_c})\gamma_5 \phi_{B_c}(x,b),
	\end{split}
\end{equation}
where $N_c=3$ is the color factor, and $b$ denotes the impact parameter conjugate to the parton transverse momentum $k_{BT}$. The distribution amplitude (DA) $\Phi_{B_c}(x,b)$ is parameterized as
\begin{equation}
	\begin{split}
	\phi_{B_c}(x,b) = \frac{f_{B_c}}{2\sqrt{2N_c}} N_{B_c} x(1-x) \exp\left[-\frac{(1-x)m_c^2 + x m_b^2}{8\beta_{B_c}^2 x(1-x)}\right] \exp\left[-2\beta_{B_c}^2 x(1-x)b^2\right],
	\end{split}
\label{eq:phibc}
\end{equation}
with $m_{b}$ and $m_{c}$ being the masses of the $b$ and $c$ quarks, respectively. The shape parameter is taken as $\beta_{B_c}=1.0\pm0.1$. The normalization constant $N_{B_{c}}$ is determined by the $B_{c}$ decay constant $f_{B_{c}}$ via the condition $\int_0^1 \phi_{B_c}(x,b=0)\, dx = \frac{f_{B_c}}{2\sqrt{2N_c}}.$

Our knowledge of the $\chi_{cJ}$ meson distribution amplitudes $\Phi_{\chi_{cJ}}$, where the dependence on the intrinsic transverse impact parameter $b$ is neglected, remains quite limited. Therefore, we adopt the models from Refs.~\cite{Liu:2018,Liu:2025}. The twist-2 distribution amplitude $\phi_{\chi_{c0}}^{v}(x)$ and the twist-3 amplitude $\phi_{\chi_{c0}}^{s}(x)$ for the scalar $\chi_{c0}$ meson are defined by
\begin{equation}
	\begin{split}
	\Phi_{\chi_{c0}}(x) = \frac{i}{\sqrt{2N_{c}}}\left\{(\not\! P\phi_{\chi_{c0}}^{v}(x)+m_{\chi_{c0}}\phi_{\chi_{c0}}^{s}(x)\right\},
	\end{split}
\end{equation}
where $P$ and $m_{\chi_{c0}}$ are the momentum and mass of the $\chi_{c0}$ meson, respectively, and $x$ is the charm quark momentum fraction. These amplitudes are assumed to take the asymptotic forms
\begin{equation}
	\begin{split}
	\begin{aligned}
		\phi_{\chi_{c0}}^{v}(x) &= 53.74\frac{f_{\chi_{c0}}}{2\sqrt{2N_{c}}}\left[x(1-x)(1-2x)\right]\mathcal{C}(x), 
	\end{aligned}
	\end{split}
\label{eq:phichic0v}
\end{equation}
\begin{equation}
	\begin{split}
		\phi_{\chi_{c0}}^{s}(x) &= 2.12\frac{f_{\chi_{c0}}}{2\sqrt{2N_{c}}}\mathcal{C}(x),
	\end{split}
\label{eq:phichic0s}
\end{equation}
with $f_{\chi_{c0}}$ denoting the decay constant. The function $\mathcal{C}(x)$ is derived from the P-wave Schr\"{o}dinger states in a Coulomb potential (see the appendix of Ref.~\cite{Liu:2018} for details):
\begin{equation}
	\begin{split}
		\mathcal{C}(x) = \left\{\left[x(1-x)(1-4x(1-x))^{3}\right]^{1/2}\Big/\left[1-4x(1-x)(1-v^{2}/4)\right]^{2}\right\}^{1-v^{2}},
	\end{split}
\end{equation}
where $v^{2}=0.3$ accounts for small relativistic corrections to the Coulomb wave functions. The distribution amplitudes $\phi_{\chi_{c0}}^{v}(x)$ and $\phi_{\chi_{c0}}^{s}(x)$ are antisymmetric and symmetric, respectively, in accordance with charge conjugation invariance for scalar mesons.

The numerical coefficients in Eqs.~\eqref{eq:phichic0v} and~\eqref{eq:phichic0s} are chosen to satisfy the normalization conditions~\cite{Liu:2018}:
\begin{equation}
	\begin{split}
	\int_{0}^{1}dx\,\phi_{\mathrm{sym}}(x) = \frac{f}{2\sqrt{2N_{c}}},
	\end{split}
\end{equation}
for the symmetric distribution amplitudes, and
\begin{equation}
	\begin{split}
	\int_{0}^{1}dx\,\phi_{\mathrm{asym}}(x)(1-2x) = \frac{f}{2\sqrt{2N_{c}}},
	\end{split}
\end{equation}
for the antisymmetric ones, where $f$ denotes a decay constant. These numerical coefficients appear because the charmonium distribution amplitudes are not expanded in Gegenbauer polynomials, unlike those for light mesons.

For the axial-vector $\chi_{c1}$ meson, the twist-2 distribution amplitude $\phi_{\chi_{c1}}^{L(T)}(x)$ and the twist-3 distribution amplitude $\phi_{\chi_{c1}}^{t(v)}(x)$ for longitudinal (transverse) polarization are defined as follows:
\begin{equation}
	\begin{split}
		\Phi_{\chi_{c1}}^{L}(x) = \frac{1}{\sqrt{2N_{c}}}\gamma_{5}\left\{m_{\chi_{c1}}\slashed{\epsilon}_{\chi_{c1}}^{L}\phi_{\chi_{c1}}^{L}(x)+\slashed{\epsilon}_{\chi_{c1}}^{L}\slashed{P}\phi_{\chi_{c1}}^{t}(x)\right\},
	\end{split}
\end{equation}
\begin{equation}
	\begin{split}
	\Phi_{\chi_{c1}}^{T}(x) = \frac{1}{\sqrt{2N_{c}}}\gamma_{5}\left\{m_{\chi_{c1}}\slashed{\epsilon}_{\chi_{c1}}^{T}\phi_{\chi_{c1}}^{v}(x)+\slashed{\epsilon}_{\chi_{c1}}^{T}\slashed{P}\phi_{\chi_{c1}}^{T}(x)\right\},
	\end{split}
\end{equation}
where $m_{\chi_{c1}}$ is the $\chi_{c1}$ mass, and $\epsilon^{L}(\epsilon^{T})$ denotes its longitudinal (transverse) polarization vector. These distribution amplitudes are expressed as
\begin{equation}
	\begin{split}
			\phi_{\chi_{c1}}^{L}(x) &= 12.60\frac{f_{\chi_{c1}}}{2\sqrt{2N_{c}}}\left[x(1-x)\right]\mathcal{C}(x), \\
	\end{split}
\end{equation}
\begin{equation}
	\begin{split}
		\phi_{\chi_{c1}}^{T}(x) &= 53.74\frac{f_{\chi_{c1}}^{\perp}}{2\sqrt{2N_{c}}}\left[x(1-x)(1-2x)\right]\mathcal{C}(x), \\
	\end{split}
\end{equation}
\begin{equation}
	\begin{split}
		\phi_{\chi_{c1}}^{t}(x) &= 23.16\frac{f_{\chi_{c1}}^{\perp}}{2\sqrt{2N_{c}}}\left[(1-2x)(1-6x+6x^{2})\right]\mathcal{C}(x), \\
	\end{split}
\end{equation}
\begin{equation}
	\begin{split}
	\phi_{\chi_{c1}}^{v}(x) &= 1.59\frac{f_{\chi_{c1}}}{2\sqrt{2N_{c}}}\left[1+(1-2x)^{2}\right]\mathcal{C}(x),\\
	\end{split}
\end{equation}
with $f_{\chi_{c1}}$ and $f_{\chi_{c1}}^{\perp}$ being the decay constants.

For P-wave two-meson distribution amplitudes (DAs) in both longitudinal and transverse polarizations, the decomposition up to twist-3 is given by~\cite{Ali:2007,Li:2017,Wang:2016,Rui:2018}:

	\begin{align}
	\Phi_{P}^{L}(z,\zeta,\omega)&= \frac{1}{\sqrt{2N_{c}}}\left[\omega\slashed{\epsilon}_{p}\phi_{P}^{0}(z,\omega^{2}) + \omega\phi_{P}^{s}(z,\omega^{2}) + \frac{\slashed{p}_{1}\slashed{p}_{2}-\slashed{p}_{2}\slashed{p}_{1}}{\omega(2\zeta-1)}\phi_{P}^{t}(z,\omega^{2})\right](2\zeta-1),\\
		\Phi_{P}^{T}(z,\zeta,\omega)&= \frac{1}{\sqrt{2N_{c}}}\left[\gamma_{5}\slashed{\epsilon}_{T}\slashed{p}\phi_{P}^{T}(z,\omega^{2}) + \omega\gamma_{5}\slashed{\epsilon}_{T}\phi_{P}^{\alpha}(z,\omega^{2}) + i\omega\frac{\epsilon^{\mu\nu\rho\sigma}\gamma_{\mu}\epsilon_{T\nu}p_{\rho}n_{-\sigma}}{p\cdot n_{-}}\phi_{P}^{\nu}(z,\omega^{2})\right]\sqrt{\zeta(1-\zeta)} + \alpha,
	\end{align}

where the kinematic parameter $\alpha$ is defined as $\alpha = \frac{(r_{1}-r_{2})^{2}}{4\eta^{2}} - \frac{r_{1}+r_{2}}{2\eta}$. The various twist components $\phi_{P}^{i}$ can be expanded in terms of Gegenbauer polynomials.

For the $\pi\pi$ and $K\pi$ system:

\begin{align}
\phi_{\pi\pi}^{0}(z,\omega^{2})&= \frac{3F_{\pi\pi}^{\parallel}(\omega^{2})}{2\sqrt{2N_{c}}}z(1-z)\left[1 + a_{2\rho}^{0}\frac{3}{2}(5t^{2}-1)\right],    \\
\phi_{\pi\pi}^{s}(z,\omega^{2})&= \frac{3F_{\pi\pi}^{\perp}(\omega^{2})}{2\sqrt{2N_{c}}}t\left[1 + a_{2\rho}^{s}(10z^{2}-10z+1)\right],     \\
\phi_{\pi\pi}^{t}(z,\omega^{2})&= \frac{3F_{\pi\pi}^{\perp}(\omega^{2})}{2\sqrt{2N_{c}}}t^{2}\left[1 + a_{2\rho}^{t}\frac{3}{2}(5t^{2}-1)\right],   \\
\phi_{\pi\pi}^{T}(z,\omega^{2})&= \frac{3F_{\pi\pi}^{\perp}(\omega^{2})}{2\sqrt{2N_{c}}}z(1-z)\left[1 + a_{2\rho}^{T}\frac{3}{2}(5t^{2}-1)\right],   \\
\phi_{\pi\pi}^{a}(z,\omega^{2})&= \frac{3F_{\pi\pi}^{\parallel}(\omega^{2})}{4\sqrt{2N_{c}}}\ t \left[1 + a_{2\rho}^{a}\,(10z^{2}-10z+1)\right],       \\
\phi_{\pi\pi}^{v}(z,\omega^{2})&=\frac{3F_{\pi\pi}^{\parallel}(\omega^{2})}{8\sqrt{2N_{c}}}\left[1 + t^{2} + a_{2\rho}^{v}\,(3t^{2}-1)\right], \\
	\phi_{K\pi}^{0}(z,\omega^{2})&=\frac{3F_{K\pi}^{\parallel}(\omega^{2})}{\sqrt{2N_{c}}}\,z(1-z)\left[1 + a_{1K^{*}}^{\parallel}3t + a_{2K^{*}}^{\parallel}\,\frac{3}{2}(5t^{2}-1)\right], \\
		\phi_{K\pi}^{s}(z,\omega^{2})&=\frac{3F_{K\pi}^{\perp}(\omega^{2})}{2\sqrt{2N_{c}}}t, \\
		\phi_{K\pi}^{t}(z,\omega^{2})&=\frac{3F_{K\pi}^{\perp}(\omega^{2})}{2\sqrt{2N_{c}}}t^{2}, \\
		\phi_{K\pi}^{T}(z,\omega^{2})&=\frac{3F_{K\pi}^{\perp}(\omega^{2})}{\sqrt{2N_{c}}}\,z(1-z)\left[1 + a_{1K^{*}}^{\perp}\,3t + a_{2K^{*}}^{\perp}\,\frac{3}{2}(5t^{2}-1)\right], \\
		\phi_{K\pi}^{a}(z,\omega^{2})&=\frac{3F_{K\pi}^{\parallel}(\omega^{2})}{4\sqrt{2N_{c}}}t, \\
		\phi_{K\pi}^{v}(z,\omega^{2})&=\frac{3F_{K\pi}^{\parallel}(\omega^{2})}{8\sqrt{2N_{c}}}(1+t^{2}),
	\end{align}
where the variable $t=1-2z$ and the Gegenbauer coefficients are taken from Refs.~\cite{Li:2017}:
\begin{equation}
	\begin{aligned}
		a_{2|\rho}^{0} &= 0.16 \pm 0.10, & a_{2|\rho}^{s} &= -0.11 \pm 0.14, & a_{2|\rho}^{t} &= -0.21 \pm 0.04, \\
		a_{2|\rho}^{T} &= 0.50 \pm 0.50, & a_{2|\rho}^{a} &= 0.40 \pm 0.40, & a_{2|\rho}^{v} &= -0.50 \pm 0.50, \\
		a_{1|K^{*}}^{\parallel} &= 0.45 \pm 0.11, & a_{2|K^{*}}^{\parallel} &= -0.75 \pm 0.08, & a_{1|K^{*}}^{\perp} &= 0.61 \pm 0.21, & a_{2|K^{*}}^{\perp} &= 0.45 \pm 0.06.
	\end{aligned}
\end{equation}
In our calculations, the twist-3 $K\pi$ distribution amplitudes $\phi_{K\pi}^{s,t}$ and $\phi_{K\pi}^{a,v}$ are taken in their asymptotic forms due to the limited available data. To gauge the uncertainty from neglecting possible non-asymptotic contributions, we consider that the first Gegenbauer moments for the twist-3 $K^*$ DAs are expected to be of similar magnitude to those in the $\rho$ system, e.g.\ $|a_{1K^*}^{t,s}| \lesssim 0.3$. Varying these moments in this range produces shifts of $\lesssim 0.5$ percentage points in $f_L$ for the $K\pi$ channels, which is well within the already quoted uncertainties from the Gegenbauer moments of the twist-2 and twist-3 $\pi\pi$ DAs.

According to the Watson theorem, elastic rescattering effects in the final-state meson pair can be absorbed into the timelike form factors $F^{\parallel,\perp}(\omega^{2})$. For the narrow intermediate resonance $K^{*}$, we parameterize the form factor $F_{K\pi}^{\parallel}(\omega^{2})$ using the relativistic Breit-Wigner (RBW) line shape~\cite{Breit:1936}:
\begin{equation}
	\begin{aligned}
	F_{K\pi}^{\parallel}(\omega^{2}) = N_{K^*}\,\frac{m_{K^{*}}^{2}}{m_{K^{*}}^{2}-\omega^{2}-i\,m_{K^{*}}\Gamma_{K^{*}}(\omega^{2})},
	\end{aligned}
\label{eq:FKpi}
\end{equation}
where the normalization factor $N_{K^*}=1.48\pm0.03$ is introduced to remedy the possible theoretical mismatch between the time-like form factor and the properties of the intermediate $K^*$ resonance, as determined in Ref.~\cite{Li:2017}. The energy-dependent width is defined as
\begin{equation}
	\begin{aligned}
	\Gamma_{K^{*}}(\omega^{2}) = \Gamma_{K^{*}}\left(\frac{m_{K^{*}}}{\omega}\right)\left(\frac{k(\omega)}{k(m_{K^{*}})}\right)^{(2L_{R}+1)},
	\end{aligned}
\end{equation}
with the orbital angular momentum in the two-meson system set to $L_{R}=1$ for a P-wave state. Here $k(\omega)$ is the momentum of a decay product in the resonance rest frame, and $k(m_{k^{*}})$ is its value at $\omega = m_{k^{*}}$. The explicit form of $k(\omega)$ is
\begin{equation}
	\begin{aligned}
		k(\omega) = \frac{\sqrt{\lambda(\omega^{2},m_{P_{1}}^{2},m_{P_{2}}^{2})}}{2\omega},
	\end{aligned}
\end{equation}
where $\lambda(a,b,c) = a^{2}+b^{2}+c^{2}-2(ab+ac+bc)$ is the K\"all\'en function.

In experimental analyses of three-body hadronic $B$ meson decays, the contribution from a broad $\rho$ resonance is often parameterized using the Gounaris-Sakurai (GS) model~\cite{Gounaris:1968}, which is based on the Breit-Wigner function. Including $\rho-\omega$ interference and contributions from excited states, the form factor $F_{\pi\pi}^{\parallel}$ is written as, with a normalization factor $N_\rho=1.05\pm0.04$ analogous to $N_{K^*}$, as determined in Ref.~\cite{Li:2017},
\begin{equation}
	\begin{aligned}
	F_{\pi\pi}^{\parallel}(\omega^{2}) = N_\rho\left[\mathrm{GS}_{\rho}(\omega^{2},m_{\rho},\Gamma_{\rho})\frac{1+c_{\omega}\,\mathrm{BW}_{\omega}(\omega^{2},m_{\omega},\Gamma_{\omega})}{1+c_{\omega}} + \sum_{j} c_{j}\,\mathrm{GS}_{j}(\omega^{2},m_{j},\Gamma_{j})\right]\left(1+\sum_{j} c_{j}\right)^{-1}
	\end{aligned}
\label{eq:Fpipi}
\end{equation}
where $m_{\rho,\omega,j}$ and $\Gamma_{\rho,\omega,j}$ (with $j=\rho^{'}(1450),\rho^{''}(1700),\rho^{'''}(2254)$) denote the masses and decay widths of the $\rho$ resonance series, as determined from $e^+e^-$ annihilation data~\cite{BaBar:2012}, and $c_{\omega}$, $c_{j}$ are the corresponding weight factors. In the present work, only the $\rho(770)$, $\rho(1450)$, and $\rho(1700)$ resonances are included in the numerical analysis; the contribution of $\rho(2254)$ is negligible in the energy region considered and is not taken into account. The GS function is given by
\begin{equation}
	\begin{aligned}
	\mathrm{GS}_{\rho}(\omega^{2},m_{\rho},\Gamma_{\rho}) = \frac{m_{\rho}^{2}\left[1+d(m_{\rho})\frac{\Gamma_{\rho}}{m_{\rho}}\right]}{m_{\rho}^{2}-\omega^{2}+f(\omega^{2},m_{\rho},\Gamma_{\rho})-i\,m_{\rho}\,\Gamma(\omega^{2},m_{\rho},\Gamma_{\rho})}
	\end{aligned}
\end{equation}
with the auxiliary functions

\begin{align}
\Gamma(s,m_{\rho},\Gamma_{\rho})&=\Gamma_{\rho}\,\frac{s}{m_{\rho}^{2}}\left(\frac{\beta_{\pi}(s)}{\beta_{\pi}(m_{\rho}^{2})}\right)^{3}, \nonumber\\
d(m)&=\frac{3m_{\pi}^{2}}{\pi g^{2}(m^{2})}\ln\left(\frac{m+2g(m^{2})}{2m_{\pi}}\right) + \frac{m}{2\pi g(m^{2})} - \frac{m_{\pi}^{2}m}{\pi g^{3}(m^{2})},   \nonumber\\
f(s,m,\Gamma)&=\frac{\Gamma m^{2}}{g^{3}(m^{2})}\left[g^{2}(s)[h(s)-h(m^{2})] + (m^{2}-s)g^{2}(m^{2})h'(m^{2})\right], \nonumber\\
g(s)&= \frac{1}{2}\sqrt{s}\,\beta_{\pi}(s), \qquad h(s) = \frac{2g(s)}{\pi \sqrt{s}}\ln\left(\frac{\sqrt{s}+2g(s)}{2m_{\pi}}\right), \qquad \beta_{\pi}(s) = \sqrt{1-\frac{4m_{\pi}^{2}}{s}}.
\end{align}
Because the form factor $F^{\perp}(\omega^{2})$ is less well determined, we estimate the ratio $\frac{F^{\parallel}(\omega^{2})}{F^{\perp}(\omega^{2})}\approx \frac{f^{T}_{V}}{f_{V}}$ using the two decay constants $f^{(T)}_{V}$ of the intermediate vector resonance. This relation is derived from vector meson dominance at $q^2 = 0$ and its application to the time-like region $\omega^2 \approx m_V^2$ is subject to theoretical uncertainties from off-shell effects and from the complex phase structure of the resonant line shapes. For the $\rho$ meson, $f_\rho^T/f_\rho \approx 0.789$; for $K^*$, $f_{K^*}^T/f_{K^*} \approx 0.853$. To estimate the sensitivity of our results to this approximation, we have varied the ratio $F^{\perp}/F^{\parallel}$ by $\pm 20\%$; the resulting shifts in the transverse polarization components $f_\parallel$ and $f_\perp$ are at the level of $1$--$2$ percentage points, which are subdominant compared to the uncertainties from $\beta_{B_c}$ and the decay constants.

\subsection{Decay amplitudes in the PQCD approach}
In this section, we compute the decay amplitudes in the PQCD approach. For the considered quasi-two-body charmed decays $B_{c}^{+}\rightarrow \Psi(R\rightarrow)P_{1}P_{2}$, the analytic formulas for the corresponding decay amplitudes can be expressed as follows:
	\begin{align}
	\mathcal{A}(B_{c}^{+}\to\chi_{c0}[\rho^{+}\to]\pi^{+}\pi^{0}) &= \frac{G_{F}}{\sqrt{2}}V_{cb}^{*}V_{ud}\left[\left(C_{2}+\frac{C_{1}}{3}\right)F_{e\chi_{c0}}^{LL}+C_{1}M_{e\chi_{c0}}^{LL}\right],\\
	\mathcal{A}(B_{c}^{+}\to\chi_{c0}[K^{*+}\to]K^{0}\pi^{+})&= \frac{G_{F}}{\sqrt{2}}V_{cb}^{*}V_{us}\left[\left(C_{2}+\frac{C_{1}}{3}\right)F_{e\chi_{c0}}^{LL}+C_{1}M_{e\chi_{c0}}^{LL}\right]. \\
	\mathcal{A}^{h}(B_{c}^{+}\to\chi_{c1}(\rho^{+}\to)\pi^{+}\pi^{0})&= \frac{G_{F}}{\sqrt{2}}V_{cb}^{*}V_{ud}\left[\left(C_{2}+\frac{C_{1}}{3}\right)F_{e\chi_{c1}}^{LL,h}+C_{1}M_{e\chi_{c1}}^{LL,h}\right], \label{eq:ampchic1rho}\\
	\mathcal{A}^{h}(B_{c}^{+}\to\chi_{c1}(K^{*+}\to)K^{0}\pi^{+})&= \frac{G_{F}}{\sqrt{2}}V_{cb}^{*}V_{us}\left[\left(C_{2}+\frac{C_{1}}{3}\right)F_{e\chi_{c1}}^{LL,h}+C_{1}M_{e\chi_{c1}}^{LL,h}\right].\label{eq:ampchic1kstar}
	\end{align}
Here, $G_{F}$ denotes the Fermi coupling constant, $V_{cb}$ and $V_{ud(s)}$ are CKM matrix elements, and $C_{1}$, $C_{2}$ are the Wilson coefficients. In Eqs.~\eqref{eq:ampchic1rho} and~\eqref{eq:ampchic1kstar}, the superscript $h=0,\parallel,\perp$ labels the longitudinal, parallel and transverse polarization components, respectively. The terms $F_{e\chi_{c0}}^{LL}$ and $F_{e\chi_{c1}}^{LL,h}$ ($M_{e\chi_{c0}}^{LL}$ and $M_{e\chi_{c1}}^{LL,h}$) arise from the factorizable (nonfactorizable) emission diagrams depicted in Fig.~\ref{fig:fig1}, corresponding to the (V-A)$\otimes$(V-A) current structure.

The explicit forms of the individual amplitudes---namely $F_{e\chi_{c0}}^{LL}$, $F_{e\chi_{c1}}^{LL,h}$, $M_{e\chi_{c0}}^{LL}$ and $M_{e\chi_{c1}}^{LL,h}$---are obtained by a direct evaluation of the Feynman diagrams in Fig.~\ref{fig:fig1}. Following the standard PQCD procedure, we arrive at the following explicit results for these amplitudes:
  	\begin{equation}
  		\begin{split}
  		F_{e\chi_{c0}}^{LL}
  			&= 8\pi C_{F}m_{B_{c}}^{4}F(\omega^{2})
  			\int_{0}^{1}\!dx_{B}\,dx_{3}
  			\int_{0}^{1/\Lambda}\!b_{B}\,db_{B}\,b_{3}\,db_{3}\,
  			\phi_{B_{c}}(x_{B},b_{B}) \\
  			&\quad\times\Big\{
  			\!-\!\big[
  			(f^{-}r(-2+r_{b}+2g^{-}x_{3})
  			-f^{+}r(-2+r_{b}+2g^{+}x_{3}))\phi^{s} \\
  			&\qquad\quad
  			+(-f^{-}g^{-}(-1+2r_{b}+g^{-}x_{3})
  			+f^{+}g^{+}(-1+2r_{b}+g^{+}x_{3}))\phi^{v}
  			\big]
  			E_{e}(t_{a})h_{a}(\alpha_{e},\beta_{a},b_{B},b_{3})S_{t}(x_{3}) \\
  			&\quad
  			+\big[
  			2r(f^{-}(g^{-}+r_{c}-x_{B})
  			-f^{+}(g^{+}+r_{c}-x_{B}))\phi^{s} \\
  			&\qquad\quad
  			+(f^{+}g^{+}(g^{-}+r_{c})
  			-f^{-}g^{-}(g^{+}+r_{c})
  			-f^{+}g^{-}x_{B}
  			+f^{-}g^{+}x_{B})\phi^{v}
  			\big]
  			h_{b}(\alpha_{e},\beta_{a},b_{3},b_{B})S_{t}(x_{B})
  			\Big\},
  		\end{split}
  	\end{equation}
   	\begin{equation}
   		\begin{split}
   		M_{e\chi_{c0}}^{LL}
   			&= \frac{-32\pi C_{F}m_{B_{c}}^{4}}{\sqrt{6}}
   			\int_{0}^{1}\!dx_{B}\,dz\,dx_{3}
   			\int_{0}^{1/\Lambda}\!b_{B}\,db_{B}\,b\,db\,
   			\phi_{B_{c}}(x_{B},b_{B})\phi_{P}^{0}(z) \\
   			&\quad\times\Big\{
   			\big[
   			r(f^{-}g^{-}x_{3}-f^{+}g^{+}x_{3}-f^{-}x_{B}+f^{+}x_{B})\phi^{s} \\
   			&\qquad\quad
   			+(f^{-}-f^{+})(-2g^{-}g^{+}x_{3}+(g^{-}+g^{+})x_{B}
   			+f^{-}g^{-}(-1+z)+f^{+}g^{+}(-1+z))\phi^{v}
   			\big]
   			E_{n}(t_{c})h_{c}(\alpha_{e},\beta_{c},b_{B},b) \\
   			&\quad
   			+\big[
   			r(-f^{-}g^{-}x_{3}+f^{+}g^{+}x_{3}+f^{-}x_{B}-f^{+}x_{B})\phi^{s} \\
   			&\qquad\quad
   			+(f^{-}g^{-}-f^{+}g^{+})((g^{-}+g^{+})x_{3}-2x_{B}+(f^{-}+f^{+})z)\phi^{v}
   			\big]
   			E_{n}(t_{d})h_{d}(\alpha_{e},\beta_{d},b_{B},b)
   			\Big\}.
   		\end{split}
   	\end{equation}
\begin{equation}
	\begin{split}
	F_{e\chi_{c1}}^{LL,0}
		&= -8\pi C_{F}m_{B_{c}}^{4}F(\omega^{2})
		\int_{0}^{1}\!dx_{B}\,dx_{3}
		\int_{0}^{1/\Lambda}\!b_{B}\,db_{B}\,b_{3}\,db_{3}\,
		\phi_{B_{c}}(x_{B},b_{B}) \\
		&\quad\times\Big\{
		\!-\!\big[
		g_{-}g_{+}\phi_{\chi_{c1}}^{t}(x_{3},b_{3})((f^{-}+f^{+})(-2+r_{b})+2(f^{-}g^{-}+f^{+}g^{+})x_{3})
		-\phi_{\chi_{c1}}^{L}(x_{3},b_{3})r_{3}(f^{-}g^{-}(-1+2r_{b} \\
		&\qquad\quad
		+g^{-}x_{3})+f^{+}g^{+}(-1+2r_{b}+g^{+}x_{3}))
		\big]
		E_{e}(t_{a})h_{a}(\alpha_{e},\beta_{a},b_{B},b_{3})S_{t}(x_{3}) \\
		&\quad
		+\phi_{\chi_{c1}}^{L}(x_{3},b_{3})\big[
		f^{+}g^{+}(g^{+}+r_{c})-f^{+}g^{+}x_{B}-f^{-}g^{+}x_{B}
		\big]
		E_{e}(t_{b})h_{b}(\alpha_{e},\beta_{b},b_{3},b_{B})S_{t}(x_{B})
		\Big\},
	\end{split}
\end{equation}
\begin{equation}
	\begin{split}
		M_{e\chi_{c1}}^{LL,0}
		&= \frac{32\pi C_{F}m_{B_{c}}^{4}}{\sqrt{6}r_{3}}
		\int_{0}^{1}\!dx_{B}\,dz\,dx_{3}
		\int_{0}^{1/\Lambda}\!b_{B}\,db_{B}\,b\,db\,
		\phi_{B_{c}}(x_{B},b_{B})\phi_{P}^{0}(z) \\
		&\quad\times\Big\{
		\big[
		g_{+}g_{-}\phi_{\chi_{c1}}^{t}(x_{3},b_{3})(f^{+}(-(g^{+}x_{3})+x_{B})
		+f^{-}(-g^{-}x_{3}+x_{B}+2f^{+}(-1+z))) \\
		&\qquad\quad
		+(f^{-}-f^{+})\phi_{\chi_{c1}}^{L}(x_{3},b_{3})
		\big] \\
		&\quad
		\times r_{3}\big[
		g^{-}x_{B}-g^{+}(x_{B}+f^{+}(-1+z))+f^{-}g^{-}(-1+z)
		\big]
		E_{n}(t_{c})h_{c}(\alpha_{e},\beta_{c},b_{B},b) \\
		&\quad
		+\big[
		(f^{-}g^{-}+f^{+}g^{+})\phi_{\chi_{c1}}^{L}(x_{3},b_{3})r_{3}((g^{-}+g^{+})x_{3}-2x_{B}+(f^{-}+f^{+})z) \\
		&\qquad\quad
		+g_{-}g_{+}\phi_{\chi_{c1}}^{t}(x_{3},b_{3})(f^{+}(-(g^{+}x_{3})+x_{B})
		+f^{-}(-(g^{-}x_{3})+x_{B}-2f^{+}z))
		\big]
		E_{n}(t_{d})h_{d}(\alpha_{e},\beta_{d},b_{B},b)
		\Big\},
	\end{split}
\end{equation}
\begin{equation}
	\begin{split}
		F_{e\chi_{c1}}^{LL,\parallel}
		&= -8\pi C_{F}m_{B_{c}}^{4}\sqrt{\eta}F(\omega^{2})
		\int_{0}^{1}\!dx_{B}\,dx_{3}
		\int_{0}^{1/\Lambda}\!b_{B}\,db_{B}\,b_{3}\,db_{3}\,
		\phi_{B_{c}}(x_{B},b_{B}) \\
		&\quad\times\Big\{
		\!-\!\big[
		(-g^{+}+g^{-}(-1+4g^{+}x_{3}))\phi_{\chi_{c1}}^{T}(x_{3},b_{3})
		-r(2+(g^{-}+g^{+})x_{3})\phi_{\chi_{c1}}^{v}(x_{3},b_{3})
		\big]
		E_{e}(t_{a})h_{a}(\alpha_{e},\beta_{a},b_{B},b_{3})S_{t}(x_{3}) \\
		&\quad
		+\phi_{\chi_{c1}}^{v}(x_{3},b_{3})r\big[
		(g^{-}+g^{+}+2r_{c}-2x_{B})
		\big]
		E_{e}(t_{b})h_{b}(\alpha_{e},\beta_{b},b_{3},b_{B})S_{t}(x_{B})
		\Big\},
	\end{split}
\end{equation}
\begin{equation}
	\begin{split}
		F_{e\chi_{c1}}^{LL,\perp}
		&= -8\pi C_{F}m_{B_{c}}^{4}\sqrt{\eta}F(\omega^{2})
		\int_{0}^{1}\!dx_{B}\,dx_{3}
		\int_{0}^{1/\Lambda}\!b_{B}\,db_{B}\,b_{3}\,db_{3}\,
		\phi_{B_{c}}(x_{B},b_{B}) \\
		&\quad\times\Big\{
		\!-\!\big[
		(g^{-}-g^{+})(\phi_{\chi_{c1}}^{T}(x_{3},b_{3})-rx_{3}\phi_{\chi_{c1}}^{v}(x_{3},b_{3}))
		\big]
		E_{e}(t_{a})h_{a}(\alpha_{e},\beta_{a},b_{B},b_{3})S_{t}(x_{3}) \\
		&\quad
		+\phi_{\chi_{c1}}^{T}(x_{3},b_{3})r(g^{+}-g^{-}) \\
		&\quad
		\times E_{e}(t_{b})h_{b}(\alpha_{e},\beta_{b},b_{3},b_{B})S_{t}(x_{B})
		\Big\},
	\end{split}
\end{equation}
\begin{equation}
	\begin{split}
		M_{e\chi_{c1}}^{LL,\parallel}
		&=-32\pi C_{F}m_{B_{c}}^{4}\sqrt{\eta}/\sqrt{6}
		\int_{0}^{1}\!dx_{B}\,dz\,dx_{3}
		\int_{0}^{1/\Lambda}\!b_{B}\,db_{B}\,b\,db\,
		\phi_{B_{c}}(x_{B},b_{B}) \\
		&\quad\times\Big\{
		\phi_{\chi_{c1}}^{T}(x_{3},b_{3})\big[
		(-2g^{-}g^{+}x_{3}+(g^{-}+g^{+})x_{B}+f^{-}g^{-}(-1+z)+f^{+}g^{+}(-1+z))\phi_{P}^{a}(z) \\
		&\qquad\quad
		+((-g^{-}+g^{+})x_{B}+f^{+}g^{+}(-1+z)+f^{-}(g^{-}-g^{-}z))\phi_{P}^{v}(z)
		\big]
		E_{n}(t_{c})h_{c}(\alpha_{e},\beta_{c},b_{B},b) \\
		&\quad
		+\big[
		((-2g^{-}g^{+}x_{3}+g^{-}x_{B}+g^{+}x_{B}-(f^{-}g^{-}+f^{+}g^{+})z)\phi^{T}(x_{3},b_{B})
		+2r((g^{-}+g^{+})x_{3}-2x_{B}+(f^{-}+f^{+})z) \\
		&\qquad\quad
		\times\phi^{v}(x_{3},b_{B}))\phi_{P}^{a}(z)
		+(-g^{-}x_{B}+g^{+}x_{B}+f^{-}g^{-}z-f^{+}g^{+}z)\phi^{T}(x_{3},b_{B})\phi_{P}^{v}(z)
		\big]
		E_{n}(t_{d})h_{d}(\alpha_{e},\beta_{d},b_{B},b)
		\Big\},
	\end{split}
\end{equation}
\begin{equation}
	\begin{split}
		M_{e\chi_{c1}}^{LL,\perp}
		&= 32\pi C_{F}m_{B_{c}}^{4}\sqrt{\eta}/\sqrt{6}
		\int_{0}^{1}\!dx_{B}\,dz\,dx_{3}
		\int_{0}^{1/\Lambda}\!b_{B}\,db_{B}\,b\,db\,
		\phi_{B_{c}}(x_{B},b_{B}) \\
		&\quad\times\Big\{
		\phi_{\chi_{c1}}^{T}(x_{3},b_{3})\big[
		(g^{-}x_{B}-g^{+}(x_{B}+f^{+}(-1+z))+f^{-}g^{-}(-1+z))\phi_{P}^{a}(z) \\
		&\qquad\quad
		-(-2g^{-}g^{+}x_{3}+(g^{-}+g^{+})x_{B}+f^{-}g^{-}(-1+z)+f^{+}g^{+}(-1+z))\phi_{P}^{v}(z)
		\big]
		E_{n}(t_{c})h_{c}(\alpha_{e},\beta_{c},b_{B},b) \\
		&\quad
		+\big[
		(g^{-}x_{B}-g^{+}x_{B}-f^{-}g^{-}z+f^{+}g^{+}z)\phi_{\chi_{c1}}^{T}(x_{3},b_{3})\phi_{P}^{v}
		+((2g^{+}g^{-}x_{3}-g^{-}x_{B}-g^{+}x_{B}+f^{-}g^{-}z+f^{+}g^{+}z)\phi^{T} \\
		&\qquad\quad
		-2r((g^{-}+g^{+})x_{3}-2x_{B}+(f^{-}+f^{+})z)\phi_{\chi_{c1}}^{v}(x_{3},b_{3})\phi_{P}^{v}
		\big]
		E_{n}(t_{d})h_{d}(\alpha_{e},\beta_{d},b_{B},b)
		\Big\},
	\end{split}
\label{eq:Mchic1perp}
\end{equation}

The color factor is taken as $C_{F}=\dfrac{4}{3}$, and the QCD scale is set to $\Lambda=0.25\pm0.05~\mathrm{GeV}$. The mass ratios are defined as $r=\frac{m_{\chi_{cJ}}}{m_{B_{c}}}$ and $r_{b(c)}=\frac{m_{b(c)}}{m_{B_{c}}}$. The explicit form of the threshold resummation factor $S_{t}(x)$ may be found in Ref.~\cite{Li:2003,Li:2005}.

In the above factorization formulas, the evolution factors $E_{e}(t)$ and $E_{n}(t)$ are given by:
\begin{equation}
	\begin{split}
		E_{e}(t) &= \alpha_{s}(t)\exp\!\big[-S_{B_{c}}(t)-S_{\Psi}(t)\big], \\
		E_{n}(t) &= \alpha_{s}(t)\exp\!\big[-S_{B_{c}}(t)-S_{V}(t)-S_{\Psi}(t)\big],
	\end{split}
\end{equation}
where the Sudakov exponents $S_{B_{c},V,\Psi}$ follow the standard PQCD formalism~\cite{Li:2003,Botts:1989,Liu:2018,Liu:2023}. The original $k_T$ factorization and Sudakov resummation were developed in Refs.~\cite{Botts:1989,Li:1992}, and the improved treatment for $B_c$ meson decays including the next-to-leading-logarithm $k_T$ resummation can be found in Refs.~\cite{Liu:2018,Liu:2020,Liu:2023}. The operator-level definition of TMD hadronic wave functions and the joint resummation are discussed in Refs.~\cite{Li:2013,Li:2014,Li:2015}:
\begin{align}
	S_{B_{c}} &= s\!\left(\frac{x_{B}m_{B_{c}}}{\sqrt{2}},b_{B}\right)
	+\frac{5}{3}\int_{m_{c}}^{t}\frac{d\bar{\mu}}{\bar{\mu}}\,
	\gamma_{q}\!\big(\alpha_{s}(\bar{\mu})\big), \label{eq:sudakovBc} \\[4pt]
	S_{V}     &= s\!\left(\frac{m_{B_{c}}}{\sqrt{2}}zf_{+},b\right)
	+s\!\left(\frac{m_{B_{c}}}{\sqrt{2}}(1-z)f_{+},b\right)
	+2\int_{1/b}^{t}\frac{d\bar{\mu}}{\bar{\mu}}\,
	\gamma_{q}\!\big(\alpha_{s}(\bar{\mu})\big), \\[4pt]
	S_{\Psi}  &= s\!\left(\frac{m_{B_{c}}}{\sqrt{2}}x_{3}g_{+},b_{3}\right)
	+s\!\left(\frac{m_{B_{c}}}{\sqrt{2}}(1-x_{3})g_{+},b_{3}\right)
	+2\int_{m_{c}}^{t}\frac{d\bar{\mu}}{\bar{\mu}}\,
	\gamma_{q}\!\big(\alpha_{s}(\bar{\mu})\big),
\end{align}
here, $\gamma_{q}=-\dfrac{\alpha_{s}}{\pi}$ denotes the quark anomalous dimension. The explicit expressions for the functions $s(Q,b)$ are provided in the Appendix of Ref.~\cite{Li:2003}.

We note that for the $B_c$ meson, the Sudakov exponent $S_{B_c}$ in Eq.~\eqref{eq:sudakovBc} is evolved from the charm quark mass $m_c$ as the infrared cutoff, rather than from $1/b$ as in the standard $B$ meson treatment. This prescription follows Refs.~\cite{Liu:2018,Liu:2020}, where it was argued that for a heavy-heavy system such as $B_c$, both the $b$ and $\bar{c}$ quarks are heavy partons whose longitudinal momenta are predominantly determined by their respective masses. The Sudakov evolution starting from $m_c$ effectively resums the double logarithms associated with the $\bar{c}$ spectator, while the $b$-quark contribution enters through the threshold resummation factor $S_t(x)$ defined earlier. This treatment has been validated in the improved PQCD formalism~\cite{Liu:2023} for $B_c \to J/\psi$ and $B_c \to \chi_{cJ}$ transitions.

The hard functions $h_{i}\,(i=a-d)$, which arise from the Fourier transformation of the virtual quark and gluon propagators, are defined as 
\begin{align}
	h_{i}(\alpha_{e},\beta_{i},b_{1},b_{2}) &= h_{1}(\alpha_{e},b_{1})h_{2}(\beta_{i},b_{1},b_{2}), \\[6pt]
	h_{1}(\alpha_{e},b_{1}) &=
	\begin{cases}
		K_{0}(\sqrt{\alpha_{e}}b_{1}), & \alpha_{e} > 0 \\[4pt]
		K_{0}(i\sqrt{-\alpha_{e}}b_{1}), & \alpha_{e} < 0
	\end{cases} \\[6pt]
	h_{2}(\beta_{i},b_{1},b_{2}) &=
	\begin{cases}
		\theta(b_{1}-b_{2})I_{0}(\sqrt{\beta_{i}}b_{2})K_{0}(\sqrt{\beta_{i}}b_{1}) + (b_{1}\leftrightarrow b_{2}), & \beta_{i} > 0 \\[4pt]
		\theta(b_{1}-b_{2})I_{0}(\sqrt{-\beta_{i}}b_{2})K_{0}(i\sqrt{-\beta_{i}}b_{1}) + (b_{1}\leftrightarrow b_{2}), & \beta_{i} < 0
	\end{cases}
\end{align}
Here, $K_{0}(ix) = \pi\big[-N_{0}(x)+iJ_{0}(x)\big]$. The virtualities $\alpha_{e}$ (for the internal gluon) and $\beta_{i}$ (for the internal quark) entering the diagrams are defined as follows:
\begin{equation}
	\begin{aligned}
		\alpha_{e} &= -(g^{-}x_{3}-x_{B})(g^{+}x_{3}-x_{B}), \\
		\beta_{a}  &= \big(r_{b}^{2}-(-1+g^{-}x_{3})(-1+g^{+}x_{3})\big), \\
		\beta_{b}  &= r_{c}^{2}+(g^{-}-x_{B})(-g^{+}+x_{B}), \\
		\beta_{c}  &= -(-g^{+}x_{3}+x_{B}+f^{-}(-1+z))(-g^{-}x_{3}+x_{B}+f^{+}(-1+z)), \\
		\beta_{d}  &= -(g^{+}x_{3}-x_{B}+f^{-}z)(g^{-}x_{3}-x_{B}+f^{+}z).
	\end{aligned}
\end{equation}

The hard scales $t_{i}$ for each diagram i=a,b,c,d are set to the maximum of the momentum transfers involved in the hard amplitudes: 
\begin{equation}
	\begin{aligned}
		t_{a,b} &= \max\!\big\{m_{B_{c}}\sqrt{|\alpha_{e}|},\,m_{B_{c}}\sqrt{|\beta_{a,b}|},\,1/b_{3},\,1/b_{B}\big\}, \\
		t_{c,d} &= \max\!\big\{m_{B_{c}}\sqrt{|\alpha_{e}|},\,m_{B_{c}}\sqrt{|\beta_{c,d}|},\,1/b,\,1/b_{B}\big\}.
	\end{aligned}
\label{eq:hardscale}
\end{equation}

\section{Numerical analysis}\label{sec:numer}

We begin this section by specifying the input parameters adopted in our numerical analysis. The masses, decay widths, CKM matrix elements, and the $B_c$ meson lifetime are taken from the Particle Data Group~\cite{PDG:2024}. The Wilson coefficients $C_i$ are initialized at the scale $\mu=m_b$ using the next-to-leading-order renormalization group evolution~\cite{Buchalla:1996}. At this scale, the relevant tree-operator Wilson coefficients take the values $C_1(m_b) = 1.082$ and $C_2(m_b) = -0.191$, evaluated in the naive dimensional regularization scheme. In the numerical evaluation of the decay amplitudes, they are evolved to the individual hard scales $t_i$ defined in Eq.~\eqref{eq:hardscale} via the renormalization group equations. The hard scale variation is taken as $0.9t \sim 1.1t$, which is the conventional choice in the standard PQCD approach for heavy-flavor decays~\cite{Li:2003,Xiao:2014}. A more conservative interval $0.75t \sim 1.25t$ could double the scale uncertainty but would not change our conclusions, as the dominant uncertainty in the present LO analysis stems from $\beta_{B_c}$ and the decay constants. The numerical values are collected as follows (in units of GeV, and the $B_c$ lifetime in ps):

\begin{align}
	m_{B_c} &= 6.275, & m_{\chi_{c0}} &= 3.415, & m_{\chi_{c1}} &= 3.511,&  m_b &= 4.8,     \nonumber \\
    m_c &= 1.5, & m_{K^\pm} &= 0.494, & m_{K^0} &= 0.498,& m_{\pi^+} &= 0.140,       \nonumber \\
	m_{\pi^0} &= 0.135, & \Gamma_\rho &= 0.1496, & \Gamma_{K^*} &= 0.0473, & \tau_{B_c} &= 0.507,
\end{align}

and the decay constants (in units of GeV), taken from Refs.~\cite{Xiao:2012,Liu:2018,Li:2017} :

\begin{align}
	f_{B_c} &= 0.489, &f_{\chi_{c0}} &= 0.093, & f_{\chi_{c1}} &= 0.185, & f_{\chi_{c1}}^\perp &= 0.090, \nonumber \\
	 f_\rho &= 0.209,& f_\rho^T &= 0.165, & f_{K^*} &= 0.217, & f_{K^*}^T &= 0.185. 
\end{align}
where the central values of $f_{\chi_{c0}}$, $f_{\chi_{c1}}$, and $f_{\chi_{c1}}^\perp$ are varied by 10\% to estimate the potential theoretical uncertainties.
\begin{align}
	f_{\chi_{c0}} = 0.093 \pm 0.009, f_{\chi_{c1}} = 0.185 \pm 0.019, f_{\chi_{c1}}^\perp = 0.090 \pm 0.009.
\end{align}
Note that the decay constants in Ref.~\cite{Beneke:1999} were originally obtained at the scale \(m_c = 1.628 \, \text{GeV}\), whereas in the present work the charm quark mass is set to \(m_c = 1.5 \, \text{GeV}\), a value commonly adopted in the standard PQCD framework. To maintain the consistency of all input parameters, we have applied the renormalization-group evolution~\cite{Beneke:2000}

\begin{equation}
	f_{c\bar{c}}(\mu) = f_{c\bar{c}}(\mu_0) \left( \frac{\alpha_s(\mu)}{\alpha_s(\mu_0)} \right)^{\gamma/b_0},
\end{equation}

with \(b_0 = 11 - 2n_f/3\), where \(n_f\) denotes the number of active quark flavors, to evolve the decay constants from \(\mu_0 = 1.628 \, \text{GeV}\) down to \(\mu = m_c = 1.5 \, \text{GeV}\). The anomalous dimensions are \(\gamma = 8C_F/3\) for \(f_{\chi_{c0}}\) and \(f_{\chi_{c2}}\) with \(C_F = 4/3\), \(\gamma = 3C_F\) for \(f_{\chi_{c1}}^\perp\) and \(f_{\chi_{c2}}^\perp\), and \(\gamma = 0\) for \(f_{\chi_{c1}}\)~\cite{Beneke:2000}. The vanishing anomalous dimension of $f_{\chi_{c1}}$ at one loop follows from the conservation of the axial-vector current in the heavy quark limit and implies that this decay constant is scale-invariant at leading logarithmic order. The numerical values quoted above have already incorporated the evolution from 1.628 GeV to 1.5 GeV.

We note that the normalization factors $N_\rho=1.05$ and $N_{K^*}=1.48$ introduced in the time-like form factors [Eqs.~\eqref{eq:FKpi} and~\eqref{eq:Fpipi}] were originally determined in the context of $B_{(s)} \to P\rho \to P\pi\pi$ decays~\cite{Li:2017} by matching the PQCD predictions to experimental data on $\rho \to \pi\pi$ and $K^* \to K\pi$ branching fractions. Their application to the $B_c$ system assumes that the mismatch between the time-like form factors and the resonance parameters is universal, i.e.\ independent of the specific heavy-quark decay that produces the resonant pair. While this universality is plausible in the PQCD framework---where the hard kernel factorizes from the two-meson DAs---it has not been tested in $B_c$ decays. The $N_{K^*}$ parameter in particular carries a $48\%$ deviation from unity, reflecting a significant departure from the naive RBW expectation. To estimate the sensitivity of our results to the assumed universality of these parameters, we have varied $N_\rho$ and $N_{K^*}$ by $\pm 20\%$ independently. The resulting shifts in the branching ratios are at the level of $^{+25\%}_{-28\%}$ for the $\pi\pi$ channels and $^{+30\%}_{-33\%}$ for the $K\pi$ channels, which are comparable to the $\beta_{B_c}$ uncertainties quoted in Tables~\ref{tab:xc0}--\ref{tab:xc1_f}. These variations are not included in the tabulated errors but should be kept in mind as an additional source of theoretical uncertainty when comparing with future experimental data.

A further source of systematic uncertainty concerns the functional form of the $B_c$ meson distribution amplitude. Throughout this work we adopt the Gaussian parametrization of Eq.~\eqref{eq:phibc}, which has been widely used in PQCD calculations~\cite{Xiao:2012,Li:2003,Liu:2018}. Alternative forms, such as the exponential model $\phi_{B_c}(x) \propto \exp[-m_{B_c}^2(1-x)^2/(2\beta_{B_c}^2)]$, can yield shape differences that are not fully captured by varying $\beta_{B_c}$ alone. In Ref.~\cite{Liu:2018}, the sensitivity of $B_c \to \chi_{cJ}$ transition form factors to the choice of $B_c$ DA model was estimated to be $\sim 10\%$--$15\%$ at the amplitude level. We therefore assign an additional $\sim 20\%$--$30\%$ uncertainty to the absolute branching ratios from this source, which does not affect the ratio $R_{\chi_{c1}/\chi_{c0}}$ where the DA model dependence largely cancels.

Based on the decay amplitudes and input parameters presented in the previous
sections, we compute the CP-averaged branching ratios of the quasi-two-body
decays $B_c^+ \to \chi_{c0,c1} [\rho(K^*) \to] \pi^+\pi^0(K^0\pi^+)$ in the
leading-order PQCD formalism. The results are summarized in
Tables~\ref{tab:xc0}--\ref{tab:nwa}. The theoretical uncertainties originate
from $\beta_{B_c}=1.0\pm0.1$, the hard scale $t=0.9t\sim1.1t$, the decay
constants $f_M$, and the Gegenbauer moments $a_\rho$ (for $\pi\pi$) and
$a_{K^*}$ (for $K\pi$). The four individual errors are added in quadrature
to obtain the total uncertainty for each observable.

\begin{table}[htbp]
	\centering
	\caption{%
		Branching ratios of $B_c^+ \to \chi_{c0} \pi^+\pi^0$ and
		$B_c^+ \to \chi_{c0} K^0\pi^+$ in the PQCD approach.
		The uncertainties are from $\beta_{B_c}$, $t$, $f_M$, and $a_\rho/a_{K^*}$, respectively.
	}
	\label{tab:xc0}
	\small
	\renewcommand{\arraystretch}{1.6}
	\begin{tabular*}{\textwidth}{@{\extracolsep{\fill}}l c c c c c@{}}
		\hline
		\hline
		\\
		Decay mode & $\mathcal{B}\,(10^{-4})$ & $\beta_{B_c}$ & $t$ & $f_M$ & $a_\rho/a_{K^*}$ \\
		\hline
		\multicolumn{6}{c}{$B_c^+ \to \chi_{c0} \pi^+\pi^0$} \\
		\\
		$[B_c^+ \to \chi_{c0}(\rho) \to \pi^+\pi^0]$
		& $7.331$ & ${}^{+\;2.278}_{-\;1.675}$ & ${}^{+\;0.107}_{-\;0.025}$ & ${}^{+\;1.488}_{-\;1.350}$ & ${}^{+\;0.040}_{-\;0.036}$ \\[4pt]
		$[B_c^+ \to \chi_{c0}(\rho(770)) \to \pi^+\pi^0]$
		& $5.806$ & ${}^{+\;1.789}_{-\;1.314}$ & ${}^{+\;0.084}_{-\;0.063}$ & ${}^{+\;1.175}_{-\;1.071}$ & ${}^{+\;0.047}_{-\;0.046}$ \\[4pt]
		$[B_c^+ \to \chi_{c0}(\rho(1450)) \to \pi^+\pi^0]$
		& $0.295$ & ${}^{+\;0.093}_{-\;0.070}$ & ${}^{+\;0.003}_{-\;0.003}$ & ${}^{+\;0.056}_{-\;0.054}$ & ${}^{+\;0.002}_{-\;0.005}$ \\[4pt]
		$[B_c^+ \to \chi_{c0}(\rho(1700)) \to \pi^+\pi^0]$
		& $0.088$ & ${}^{+\;0.031}_{-\;0.022}$ & ${}^{+\;0.001}_{-\;0.001}$ & ${}^{+\;0.018}_{-\;0.016}$ & ${}^{+\;0.001}_{-\;0.001}$ \\
		\\
		\hline
		\multicolumn{6}{c}{$B_c^+ \to \chi_{c0} K^0 \pi^+$} \\
		\\
		$[B_c^+ \to \chi_{c0}(K^*(892)) \to K^0 \pi^+]$
		& $1.487$ & ${}^{+\;0.464}_{-\;0.342}$ & ${}^{+\;0.021}_{-\;0.017}$ & ${}^{+\;0.302}_{-\;0.274}$ & ${}^{+\;0.003}_{-\;0.004}$ \\
		\\
		\hline
		\hline
	\end{tabular*}
\end{table}

\begin{figure}[htbp]
	\centering
	\includegraphics[width=0.5\textwidth]{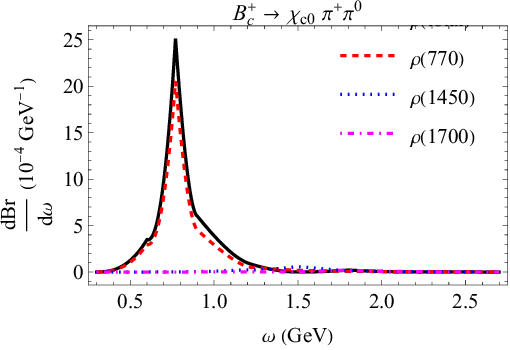}
	\caption{Differential branching ratio $d\mathcal{B}/d\omega$ (in units of
	$10^{-4}\ \mathrm{GeV}^{-1}$) of $B_c^+\to\chi_{c0}\pi^+\pi^0$ as a function
	of the $\pi\pi$ invariant mass $\omega$. The solid, dashed, dotted, and
	dash-dotted curves correspond to the coherent sum of $\rho(770)$,
	$\rho(1450)$, and $\rho(1700)$, and to the individual $\rho(770)$,
	$\rho(1450)$, and $\rho(1700)$ contributions, respectively.}
	\label{fig:chi0_rho}
\end{figure}

From Table~\ref{tab:xc0}, the coherent sum of $\rho(770)$, $\rho(1450)$,
and $\rho(1700)$ yields $\mathcal{B}(B_c^+ \to \chi_{c0} \pi^+\pi^0) =
7.33^{+2.28}_{-1.68} \times 10^{-4}$ ($0.733 \times 10^{-3}$). The
$\rho(770)$ resonance dominates with $\mathcal{B} = 5.81^{+1.79}_{-1.31}
\times 10^{-4}$ ($0.581 \times 10^{-3}$), approximately $79\%$ of the total.
The coherent sum exceeds the single $\rho(770)$ contribution by about $26\%$,
a consequence of the complex coefficients $c_1=0.158\,e^{i3.76}$ and
$c_2=0.068\,e^{i1.39}$ in the GS model, whose relative phases lead to
constructive interference in the low invariant mass region. The $\rho(1450)$ and $\rho(1700)$
contributions are $0.295\times10^{-4}$ and $0.088\times10^{-4}$, respectively. The
combined excited-state contribution is thus about $5.2\%$ of the total. The
$\pi\pi$ invariant-mass dependence of the differential branching ratio
$d\mathcal{B}/d\omega$ is displayed in Fig.~\ref{fig:chi0_rho}, where the
$\rho(770)$ resonance appears as a sharp peak at $\omega\simeq 0.77$ GeV,
while the $\rho(1450)$ and $\rho(1700)$ excitations produce much smaller
enhancements at larger $\omega$.

\begin{figure}[htbp]
	\centering
	\includegraphics[width=0.5\textwidth]{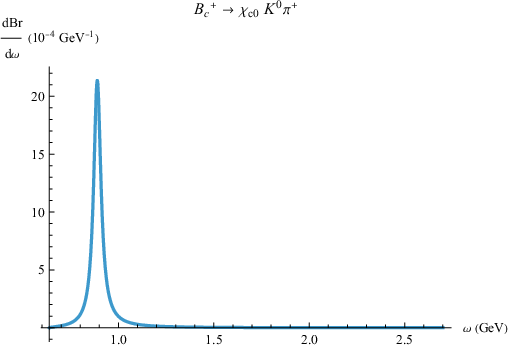}
	\caption{Differential branching ratio $d\mathcal{B}/d\omega$ (in units of
	$10^{-4}\ \mathrm{GeV}^{-1}$) of $B_c^+\to\chi_{c0}K^0\pi^+$ as a function
	of the $K\pi$ invariant mass $\omega$. The narrow $K^*(892)$ resonance
	produces a single sharp peak.}
	\label{fig:chi0_Kpi}
\end{figure}

For $B_c^+ \to \chi_{c0} K^0\pi^+$ via $K^*(892)$, the branching ratio is
$1.49 \times 10^{-4}$. Under isospin symmetry, $\mathcal{B}(B_c^+ \to
\chi_{c0} K^+\pi^0) = \frac{1}{2}\mathcal{B}(B_c^+ \to \chi_{c0} K^0\pi^+)
\approx 0.74 \times 10^{-4}$. The corresponding $K\pi$ invariant-mass
distribution is displayed in Fig.~\ref{fig:chi0_Kpi}, where the narrow
$K^*(892)$ resonance produces a single sharp peak at $\omega\simeq 0.89$ GeV.

\begin{table}[htbp]
	\centering
	\caption{%
		Branching ratios of $B_c^+ \to \chi_{c1} \pi^+\pi^0$ and
		$B_c^+ \to \chi_{c1} K^0\pi^+$ in the PQCD approach.
	}
	\label{tab:xc1}
	\small
	\renewcommand{\arraystretch}{1.6}
	\begin{tabular*}{\textwidth}{@{\extracolsep{\fill}}l c c c c c@{}}
		\hline
		\hline
		\\
		Decay mode & $\mathcal{B}\,(10^{-4})$ & $\beta_{B_c}$ & $t$ & $f_M$ & $a_\rho/a_{K^*}$ \\
		\hline
		\multicolumn{6}{c}{$B_c^+ \to \chi_{c1} \pi^+\pi^0$} \\
		\\
		$[B_c^+ \to \chi_{c1}(\rho) \to \pi^+\pi^0]$
		& $10.153$ & ${}^{+\;3.917}_{-\;2.568}$ & ${}^{+\;0.244}_{-\;0.056}$ & ${}^{+\;2.204}_{-\;1.989}$ & ${}^{+\;0.120}_{-\;0.000}$ \\[4pt]
		$[B_c^+ \to \chi_{c1}(\rho(770)) \to \pi^+\pi^0]$
		& $7.830$ & ${}^{+\;2.974}_{-\;2.051}$ & ${}^{+\;0.119}_{-\;0.110}$ & ${}^{+\;1.700}_{-\;1.535}$ & ${}^{+\;0.003}_{-\;0.000}$ \\[4pt]
		$[B_c^+ \to \chi_{c1}(\rho(1450)) \to \pi^+\pi^0]$
		& $0.441$ & ${}^{+\;0.170}_{-\;0.116}$ & ${}^{+\;0.005}_{-\;0.007}$ & ${}^{+\;0.096}_{-\;0.086}$ & ${}^{+\;0.011}_{-\;0.009}$ \\[4pt]
		$[B_c^+ \to \chi_{c1}(\rho(1700)) \to \pi^+\pi^0]$
		& $0.160$ & ${}^{+\;0.062}_{-\;0.042}$ & ${}^{+\;0.001}_{-\;0.002}$ & ${}^{+\;0.035}_{-\;0.031}$ & ${}^{+\;0.007}_{-\;0.006}$ \\
		\\
		\hline
		\multicolumn{6}{c}{$B_c^+ \to \chi_{c1} K^0 \pi^+$} \\
		\\
		$[B_c^+ \to \chi_{c1}(K^*(892)) \to K^0 \pi^+]$
		& $1.940$ & ${}^{+\;0.713}_{-\;0.494}$ & ${}^{+\;0.033}_{-\;0.024}$ & ${}^{+\;0.419}_{-\;0.367}$ & ${}^{+\;0.008}_{-\;0.003}$ \\
		\\
		\hline
		\hline
	\end{tabular*}
\end{table}

\begin{figure}[htbp]
	\centering
	\includegraphics[width=0.5\textwidth]{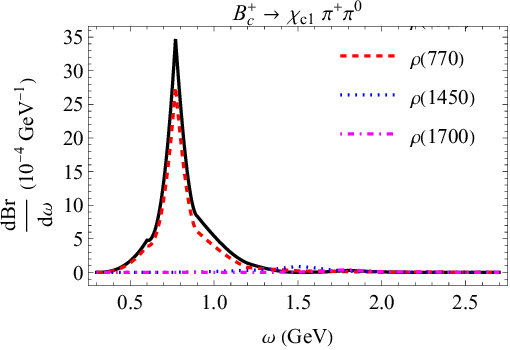}
	\caption{Differential branching ratio $d\mathcal{B}/d\omega$ (in units of
	$10^{-4}\ \mathrm{GeV}^{-1}$) of $B_c^+\to\chi_{c1}\pi^+\pi^0$ as a function
	of the $\pi\pi$ invariant mass $\omega$. The solid, dashed, dotted, and
	dash-dotted curves correspond to the coherent sum of $\rho(770)$,
	$\rho(1450)$, and $\rho(1700)$, and to the individual $\rho(770)$,
	$\rho(1450)$, and $\rho(1700)$ contributions, respectively.}
	\label{fig:chi1_rho}
\end{figure}

For $B_c^+ \to \chi_{c1} \pi^+\pi^0$ (Table~\ref{tab:xc1}), the total
branching ratio from the coherent $\rho$ sum is $\mathcal{B} =
10.15^{+3.92}_{-2.57} \times 10^{-4}$ ($1.02 \times 10^{-3}$), about
$38\%$ larger than the $\chi_{c0}$ counterpart, reflecting the larger
$f_{\chi_{c1}}$ and the additional transverse polarization contributions.
The $\rho(770)$ contributes $\mathcal{B} = 7.83^{+2.97}_{-2.05} \times
10^{-4}$ ($0.783 \times 10^{-3}$). For the $K\pi$ channel,
$\mathcal{B}(B_c^+ \to \chi_{c1} K^0\pi^+) = 1.94 \times 10^{-4}$,
with $K^+\pi^0$ receiving half of this value. The corresponding distribution
for $B_c^+\to\chi_{c1}\pi^+\pi^0$ is shown in Fig.~\ref{fig:chi1_rho}; its
shape is similar, with the $\rho(770)$ peak again dominating the spectrum.

\begin{figure}[htbp]
	\centering
	\includegraphics[width=0.5\textwidth]{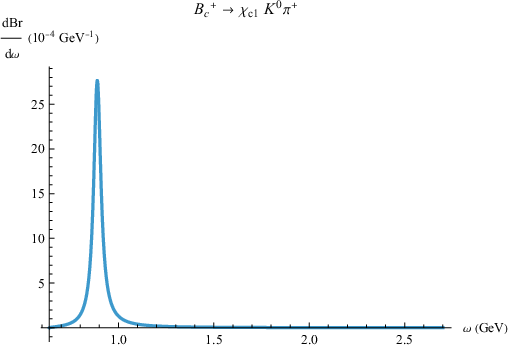}
	\caption{Differential branching ratio $d\mathcal{B}/d\omega$ (in units of
	$10^{-4}\ \mathrm{GeV}^{-1}$) of $B_c^+\to\chi_{c1}K^0\pi^+$ as a function
	of the $K\pi$ invariant mass $\omega$. The narrow $K^*(892)$ resonance
	produces a single sharp peak.}
	\label{fig:chi1_Kpi}
\end{figure}

The $K\pi$ distribution for $B_c^+ \to \chi_{c1} K^0\pi^+$ is displayed in
Fig.~\ref{fig:chi1_Kpi}; as in the $\chi_{c0}$ case it is dominated by a
single narrow $K^*(892)$ peak.

\begin{table}[htbp]
	\centering
	\caption{%
		Polarization fractions $f_i=\mathcal{B}_i/\mathcal{B}_{\rm total}$ (in \%)
		for $B_c^+ \to \chi_{c1} \pi^+\pi^0$ and $B_c^+ \to \chi_{c1} K^0\pi^+$
		in the PQCD approach.
	}
	\label{tab:xc1_f}
	\small
	\renewcommand{\arraystretch}{1.6}
	\begin{tabular*}{\textwidth}{@{\extracolsep{\fill}}l c c c c c@{}}
		\hline
		\hline
		\\
		Decay mode & $f$ (\%) & $\beta_{B_c}$ & $t$ & $f_M$ & $a_\rho/a_{K^*}$ \\
		\hline
		\multicolumn{6}{c}{$B_c^+ \to \chi_{c1} \pi^+\pi^0$ \quad $f_L$} \\
		\\
		$[B_c^+ \to \chi_{c1}(\rho) \to \pi^+\pi^0]$
		& $93.31$ & ${}^{+\;0.24}_{-\;0.46}$ & ${}^{+\;0.00}_{-\;0.13}$ & ${}^{+\;0.05}_{-\;0.02}$ & ${}^{+\;0.00}_{-\;0.12}$ \\[4pt]
		$[B_c^+ \to \chi_{c1}(\rho(770)) \to \pi^+\pi^0]$
		& $93.25$ & ${}^{+\;0.40}_{-\;0.35}$ & ${}^{+\;0.02}_{-\;0.00}$ & ${}^{+\;0.03}_{-\;0.00}$ & ${}^{+\;0.12}_{-\;0.12}$ \\[4pt]
		$[B_c^+ \to \chi_{c1}(\rho(1450)) \to \pi^+\pi^0]$
		& $75.48$ & ${}^{+\;1.19}_{-\;1.04}$ & ${}^{+\;0.00}_{-\;0.09}$ & ${}^{+\;0.07}_{-\;0.07}$ & ${}^{+\;0.18}_{-\;0.08}$ \\[4pt]
		$[B_c^+ \to \chi_{c1}(\rho(1700)) \to \pi^+\pi^0]$
		& $63.98$ & ${}^{+\;1.54}_{-\;1.30}$ & ${}^{+\;0.00}_{-\;0.17}$ & ${}^{+\;0.10}_{-\;0.07}$ & ${}^{+\;0.95}_{-\;0.74}$ \\
		\\
		\hline
		\multicolumn{6}{c}{$B_c^+ \to \chi_{c1} \pi^+\pi^0$ \quad $f_\parallel$} \\
		\\
		$[B_c^+ \to \chi_{c1}(\rho) \to \pi^+\pi^0]$
		& $6.01$ & ${}^{+\;0.44}_{-\;0.21}$ & ${}^{+\;0.14}_{-\;0.00}$ & ${}^{+\;0.02}_{-\;0.04}$ & ${}^{+\;0.06}_{-\;0.00}$ \\[4pt]
		$[B_c^+ \to \chi_{c1}(\rho(770)) \to \pi^+\pi^0]$
		& $6.05$ & ${}^{+\;0.34}_{-\;0.36}$ & ${}^{+\;0.01}_{-\;0.00}$ & ${}^{+\;0.02}_{-\;0.01}$ & ${}^{+\;0.06}_{-\;0.06}$ \\[4pt]
		$[B_c^+ \to \chi_{c1}(\rho(1450)) \to \pi^+\pi^0]$
		& $22.58$ & ${}^{+\;1.01}_{-\;1.14}$ & ${}^{+\;0.09}_{-\;0.00}$ & ${}^{+\;0.06}_{-\;0.06}$ & ${}^{+\;0.23}_{-\;0.32}$ \\[4pt]
		$[B_c^+ \to \chi_{c1}(\rho(1700)) \to \pi^+\pi^0]$
		& $33.62$ & ${}^{+\;1.28}_{-\;1.51}$ & ${}^{+\;0.15}_{-\;0.00}$ & ${}^{+\;0.07}_{-\;0.09}$ & ${}^{+\;0.89}_{-\;1.11}$ \\
		\\
		\hline
		\multicolumn{6}{c}{$B_c^+ \to \chi_{c1} \pi^+\pi^0$ \quad $f_\perp$} \\
		\\
		$[B_c^+ \to \chi_{c1}(\rho) \to \pi^+\pi^0]$
		& $0.68$ & ${}^{+\;0.01}_{-\;0.03}$ & ${}^{+\;0.00}_{-\;0.01}$ & ${}^{+\;0.00}_{-\;0.00}$ & ${}^{+\;0.06}_{-\;0.06}$ \\[4pt]
		$[B_c^+ \to \chi_{c1}(\rho(770)) \to \pi^+\pi^0]$
		& $0.70$ & ${}^{+\;0.00}_{-\;0.04}$ & ${}^{+\;0.00}_{-\;0.02}$ & ${}^{+\;0.00}_{-\;0.02}$ & ${}^{+\;0.06}_{-\;0.06}$ \\[4pt]
		$[B_c^+ \to \chi_{c1}(\rho(1450)) \to \pi^+\pi^0]$
		& $1.93$ & ${}^{+\;0.03}_{-\;0.05}$ & ${}^{+\;0.01}_{-\;0.00}$ & ${}^{+\;0.01}_{-\;0.01}$ & ${}^{+\;0.14}_{-\;0.15}$ \\[4pt]
		$[B_c^+ \to \chi_{c1}(\rho(1700)) \to \pi^+\pi^0]$
		& $2.40$ & ${}^{+\;0.02}_{-\;0.03}$ & ${}^{+\;0.02}_{-\;0.00}$ & ${}^{+\;0.01}_{-\;0.01}$ & ${}^{+\;0.16}_{-\;0.16}$ \\
		\\
		\hline
		\multicolumn{6}{c}{$B_c^+ \to \chi_{c1} K^0 \pi^+$} \\
		\\
		$[B_c^+ \to \chi_{c1}(K^*(892)) \to K^0 \pi^+] \quad f_L$
		& $91.97$ & ${}^{+\;0.48}_{-\;0.45}$ & ${}^{+\;0.02}_{-\;0.00}$ & ${}^{+\;0.00}_{-\;0.26}$ & ${}^{+\;0.03}_{-\;0.02}$ \\[4pt]
		$[B_c^+ \to \chi_{c1}(K^*(892)) \to K^0 \pi^+] \quad f_\parallel$
		& $7.23$ & ${}^{+\;0.42}_{-\;0.45}$ & ${}^{+\;0.00}_{-\;0.02}$ & ${}^{+\;0.24}_{-\;0.00}$ & ${}^{+\;0.01}_{-\;0.03}$ \\[4pt]
		$[B_c^+ \to \chi_{c1}(K^*(892)) \to K^0 \pi^+] \quad f_\perp$
		& $0.80$ & ${}^{+\;0.03}_{-\;0.03}$ & ${}^{+\;0.00}_{-\;0.00}$ & ${}^{+\;0.02}_{-\;0.00}$ & ${}^{+\;0.00}_{-\;0.00}$ \\
		\\
		\hline
		\hline
	\end{tabular*}
\end{table}

The polarization fractions are listed in Table~\ref{tab:xc1_f}. For the
$\rho(770)$-dominated $\pi\pi$ channels, $f_L = 93.3\%$ dominates, a generic PQCD feature: factorizable emission diagrams produce longitudinal
$\rho$ polarization at leading power in $1/m_b$, while transverse amplitudes
are power-suppressed by $\Lambda_{\rm QCD}/m_b$. This is consistent with
$f_0 \approx 90\%$ found in
$B_c \to J/\psi \pi\pi$~\cite{Wang:2026}. The individual polarization
components for the coherent $\rho$ sum are $\mathcal{B}_L = 9.47 \times
10^{-4}$, $\mathcal{B}_\parallel = 0.61 \times 10^{-4}$,
$\mathcal{B}_\perp = 0.069 \times 10^{-4}$; for the single $\rho(770)$,
$\mathcal{B}_L = 7.30 \times 10^{-4}$,
$\mathcal{B}_\parallel = 0.474 \times 10^{-4}$,
$\mathcal{B}_\perp = 0.054 \times 10^{-4}$.

As the $\rho$ resonance mass increases, $f_L$ decreases from $93.3\%$
($\rho(770)$) to $75.5\%$ ($\rho(1450)$) and $64.0\%$ ($\rho(1700)$),
while $f_\parallel$ grows from $6.0\%$ to $22.6\%$ and $33.6\%$.
$f_\perp$ remains below $2.4\%$ throughout. This trend reflects the
$\chi_{c1}$ amplitude structure: $\mathcal{A}_0$ is dominated by the
twist-2 DA $\phi_{\chi_{c1}}^L$, while $\mathcal{A}_\parallel$ receives
both twist-2 and twist-3 contributions whose relative weight grows
with the $\pi\pi$ invariant mass $\omega$~\cite{Liu:2018}. For the
$K\pi$ channel, $f_L = 92.0\%$, within $2\%$ of the $\pi\pi$ value,
due to the non-zero $a_{1K^*}^\parallel = 0.45$~\cite{Li:2017} and the ratio
$f_{K^*}^T/f_{K^*} \approx 0.853$ versus $f_\rho^T/f_\rho \approx
0.789$. The total uncertainties on the polarization fractions are
$1$--$2$ percentage points, smaller than those of the branching ratios
since correlated variations largely cancel.

The smallness of $f_\perp \approx 0.7\%$ for the $\rho(770)$ channel
deserves attention. Inspecting the perpendicular amplitude
$\mathcal{M}_{e\chi_{c1}}^{LL,\perp}$ in Eq.~\eqref{eq:Mchic1perp}, one observes that the
$E_n(t_c)$ and $E_n(t_d)$ terms each receive contributions proportional
to $\phi_{\chi_{c1}}^T \phi_P^a$ and $\phi_{\chi_{c1}}^T \phi_P^v$, with
the twist-3 DA $\phi_{\chi_{c1}}^v$ also entering through the
nonfactorizable diagrams. The relative signs between the $h_c$ and $h_d$
terms induce a partial cancellation, and the overall magnitude is further
suppressed by the small overlap of the antisymmetric $\phi_{\chi_{c1}}^T(x)$
with the two-meson DAs in the integrand. A measurement of $f_\perp$ would provide a test of the factorization formalism.

\begin{table}[htbp]
	\centering
	\caption{%
		Two-body branching fractions $\mathcal{B}(B_c^+ \to \chi_{cJ} V)$
		estimated from the quasi-two-body results under the narrow-width approximation
		$\mathcal{B}(B_c \to \chi_{cJ} V) \simeq \mathcal{B}(B_c \to \chi_{cJ}
		[V \to] P_1P_2) / \mathcal{B}(V \to P_1P_2)$, with
		$\mathcal{B}(\rho \to \pi\pi) \approx 100\%$ and
		$\mathcal{B}(K^* \to K\pi) \approx 100\%$~\cite{PDG:2024}.
		The total errors are obtained by adding the four individual uncertainties
		in quadrature. For reference, predictions from two-body approaches
			are also shown; these treat the $V$ meson as an on-shell final-state
			particle and are not directly equivalent to the NWA-extracted values.
	}
	\label{tab:nwa}
	\small
	\renewcommand{\arraystretch}{1.6}
	\begin{tabular*}{\textwidth}{@{\extracolsep{\fill}}l c c c@{}}
		\hline
		\hline
		\\
		Decay mode & This work & Ref.~\cite{Liu:2025} (iPQCD) & Two-body approaches (for reference) \\
		\hline
		$B_c^+ \to \chi_{c0} \rho^+$
		& $7.33^{+2.72}_{-2.15} \times 10^{-4}$ & $2.97 \times 10^{-4}$ & $5.8^{\rm a}\times10^{-3}$, $1.69^{\rm b}\times10^{-3}$, $1.3^{\rm c}\times10^{-3}$, $0.58^{\rm d}\times10^{-3}$ \\[4pt]
		$B_c^+ \to \chi_{c1} \rho^+$
		& $10.15^{+4.50}_{-3.25} \times 10^{-4}$ & $1.40 \times 10^{-3}$ & $2.8^{\rm a}\times10^{-3}$, $0.43^{\rm b}\times10^{-3}$, $0.29^{\rm c}\times10^{-3}$, $0.15^{\rm d}\times10^{-3}$ \\[4pt]
		$B_c^+ \to \chi_{c0} K^{*+}$
		& $1.49^{+0.55}_{-0.44} \times 10^{-4}$ & $2.00 \times 10^{-5}$ & $96^{\rm b}\times10^{-6}$, $70^{\rm c}\times10^{-6}$, $40^{\rm d}\times10^{-6}$, $330^{\rm a}\times10^{-6}$ \\[4pt]
		$B_c^+ \to \chi_{c1} K^{*+}$
		& $1.94^{+0.83}_{-0.62} \times 10^{-4}$ & $2.49 \times 10^{-5}$ & $27^{\rm b}\times10^{-6}$, $18^{\rm c}\times10^{-6}$, $10^{\rm d}\times10^{-6}$, $180^{\rm a}\times10^{-6}$ \\
		\\
		\hline
		\hline
	\end{tabular*}
	\vspace{2pt}
	{\footnotesize $^{\rm a}$Ref.~\cite{Rui:2018b} (PQCD);\ \ $^{\rm b}$Ref.~\cite{Zhang:2023} (covariant light-front);\ \ $^{\rm c}$Ref.~\cite{Ivanov:2006} (covariant light-front);\ \ $^{\rm d}$Ref.~\cite{Ebert:2010} (relativistic quark model).}
\end{table}

Several relative ratios provide insight into the underlying dynamics. We
consider two complementary observables: the SU(3)-breaking ratio $R_{K/\pi}$
between the Cabibbo-suppressed $K\pi$ and Cabibbo-favored $\pi\pi$ channels,
which probes the light-meson side, and the spin ratio
$R_{\chi_{c1}/\chi_{c0}}$ between the $\chi_{c1}$ and $\chi_{c0}$ yields,
which probes the charmonium side.

For SU(3) breaking, since the $\chi_{cJ}$ side is common to both channels,
the ratio isolates the flavor-breaking structure of the light-meson system
and the CKM hierarchy. Its value follows from the naive-factorization estimate,
\begin{equation}
R_{K/\pi}^{\rm naive} \simeq
\left|\frac{V_{us}}{V_{ud}}\right|^2 \times
\left(\frac{f_{K^*}}{f_\rho}\right)^2 \times
\left(\frac{N_{K^*}}{N_\rho}\right)^2 \times
\frac{\Phi(K\pi)}{\Phi(\pi\pi)}
\approx 0.053 \to 0.057 \to 0.114 \to 0.40,
\end{equation}
where the successive factors account for the CKM suppression, the decay
constants, the normalization parameters, and the phase space. With the
standard K\"all\'en momentum in the energy-dependent width, the narrow
$K^*(892)$ resonance makes this last factor an \emph{enhancement} of
$\Phi(K\pi)/\Phi(\pi\pi)\approx 3.5$, the narrow-width ratio of the
resonance integrals $\pi m_{K^*}^2/(2\Gamma_{K^*})/[\pi
m_\rho^2/(2\Gamma_\rho)]\approx 4.2$ being reduced by the $K\pi$ versus
$\pi\pi$ phase-space factors (the value drops to $\approx 3.1$ if the
Gounaris--Sakurai $\rho$ line shape is used instead of a Breit--Wigner).
The resulting naive estimate $R_{K/\pi}^{\rm naive}\approx 0.40$ is roughly
a factor of two above the full PQCD results,
\begin{equation}
R_{K/\pi}^{\chi_{c0}} \equiv \frac{\mathcal{B}(B_c^+ \to \chi_{c0} K^0\pi^+)}
{\mathcal{B}(B_c^+ \to \chi_{c0} \pi^+\pi^0)} \approx 0.20,\qquad
R_{K/\pi}^{\chi_{c1}} \equiv \frac{\mathcal{B}(B_c^+ \to \chi_{c1} K^0\pi^+)}
{\mathcal{B}(B_c^+ \to \chi_{c1} \pi^+\pi^0)} \approx 0.19,
\end{equation}
the deficit reflecting the genuine SU(3) breaking in the reduced amplitudes
beyond the factorized factors---in particular the non-vanishing Gegenbauer
moment $a_{1K^*}^{\parallel}=0.45$ of the $K^*$ two-meson distribution
amplitude, which has no $\rho$ counterpart. The two ratios are mutually
consistent and essentially independent of the charmonium spin state, as
expected for a light-meson-side effect. The normalization-parameter ratio
$N_{K^*}/N_\rho \approx 1.41$ therefore provides only a partial compensation
of the CKM suppression.
This SU(3)-breaking ratio is the direct $\chi_{cJ}$-sector analogue of the
ratio $R_{K/\pi}$ that LHCb has already measured in the $J/\psi$ sector
~\cite{LHCb:2016}, making it an experimentally accessible cross-check of the
light-meson dynamics.

The spin ratio between the $\chi_{c1}$ and $\chi_{c0}$ total branching
fractions,
\begin{equation}
R_{\chi_{c1}/\chi_{c0}}^{\pi\pi} \equiv
\frac{\mathcal{B}(B_c^+ \to \chi_{c1} \pi^+\pi^0)}
     {\mathcal{B}(B_c^+ \to \chi_{c0} \pi^+\pi^0)} \approx 1.38,
\end{equation}
is a key prediction of this work. In the quasi-two-body framework both
$\chi_{c0}$ and $\chi_{c1}$ are produced through the same factorizable
emission topology via the intermediate $\rho$ resonance, so their relative
yield is governed by a competition between the charmonium decay-constant
hierarchy, which favors $\chi_{c1}$, and phase space, which favors
$\chi_{c0}$. The decay-constant hierarchy alone would suggest
$R_{\chi_{c1}/\chi_{c0}}^{\rm naive}\sim(f_{\chi_{c1}}/f_{\chi_{c0}})^2\sim 4$;
the inclusion of phase space pulls this down to $\approx 1.38$. A measurement
of $R_{\chi_{c1}/\chi_{c0}}^{\pi\pi}$ would therefore discriminate between
the resonant quasi-two-body production mechanism, in which $\chi_{c0}$ and
$\chi_{c1}$ share the same production topology, and the direct two-body production of
the charmonium pair.

This spin ratio is strongly channel dependent. In the direct two-body
transition to a light pseudoscalar, $B_c^+\to\chi_{cJ}\pi^+$, the
$\chi_{c1}$ state is the most strongly suppressed of the three P-wave
charmonia: Ref.~\cite{Liu:2025} obtains
$R_{\chi_{c1}/J/\psi}=3.56\times10^{-2}$ and
$R_{\chi_{c0}/J/\psi}=0.17$, i.e.\ $R_{\chi_{c1}/\chi_{c0}}\approx0.21$,
consistent with the LHCb bound
$B(\chi_{c1}\pi^+)/B(\chi_{c2}\pi^+)<0.49$~\cite{LHCb:2024}. In the
direct two-body vector channel $B_c^+\to\chi_{cJ}\rho^+$, by contrast,
Ref.~\cite{Liu:2025} finds $R_{\chi_{c1}/\chi_{c0}}\approx4.7$, owing to
a constructive twist-2--twist-3 interference that is destructive in the
pseudoscalar channel. Our quasi-two-body value
$R_{\chi_{c1}/\chi_{c0}}^{\pi\pi}\approx1.38$ lies between these two
two-body limits, reflecting the distinct structure of the off-shell
resonant $\rho$ encoded in the two-meson distribution amplitude.

The analogous ratio for the Cabibbo-suppressed $K\pi$ channel,
\begin{equation}
R_{\chi_{c1}/\chi_{c0}}^{K\pi} \equiv
\frac{\mathcal{B}(B_c^+ \to \chi_{c1} K^0\pi^+)}
     {\mathcal{B}(B_c^+ \to \chi_{c0} K^0\pi^+)} \approx 1.30,
\end{equation}
is a further prediction of the same production mechanism. Through the
	$K^*(892)$ resonance, $\chi_{c0}$ and $\chi_{c1}$ are again produced via
	the identical factorizable emission topology, so the ratio is governed by
	the same charmonium-side dynamics as in the $\pi\pi$ case. The near-equality of
	$R_{\chi_{c1}/\chi_{c0}}^{K\pi} \approx 1.30$ to
	$R_{\chi_{c1}/\chi_{c0}}^{\pi\pi} \approx 1.38$ confirms that the ratio is
	dominated by the charmonium-side dynamics, with only a mild residual
	sensitivity to the light-meson distribution amplitudes. The remaining
	difference reflects the distinct structure of the $K^*$ and $\rho$
	mesons---notably the non-vanishing Gegenbauer moment
	$a_{1K^*}^{\parallel}=0.45$~\cite{Li:2017}, which vanishes for the $\rho$---
	and indicates that light-meson DA effects are subdominant in this
	observable.

For reference, Table~\ref{tab:compare} lists our NWA-extracted two-body
branching fractions alongside the direct two-body iPQCD results of
Ref.~\cite{Liu:2025}. The numerical values differ significantly---the
$\chi_{c0}\rho^+$ channel, for instance, is $7.33\times10^{-4}$ in our
NWA extraction versus $2.97\times10^{-4}$ in the two-body iPQCD calculation.
Several factors contribute to this difference. First, the two calculations
operate in different factorization frameworks: the quasi-two-body approach
integrates over the $\pi\pi$ invariant mass $\omega^2$, while the two-body
calculation treats the $\rho$ as an on-shell final-state particle. The
off-shell dynamics and the $\omega^2$ integration inherent in the
quasi-two-body formulation naturally modify the convolution of the hard
kernel with the meson distribution amplitudes. Second, the present work
is performed at leading order in the PQCD expansion, whereas
Ref.~\cite{Liu:2025} employs the improved PQCD (iPQCD) framework, also
at leading order in $\alpha_s$ but with next-to-leading-logarithm $k_T$
resummation~\cite{Liu:2020} and an improved treatment of charm-quark
effects. The relativistic corrections to the heavy-quarkonium light-cone
distribution amplitudes have been shown to be comparable with the
next-to-leading-order radiative corrections~\cite{Wang:2017}. A quantitative
decomposition of these
two sources of difference requires a quasi-two-body calculation at the
iPQCD level, which is beyond the scope of the present work.

We emphasize, however, that the ratio $R_{\chi_{c1}/\chi_{c0}}$ is less
sensitive to the absolute normalization uncertainties discussed above,
since the $\chi_{c0}$ and $\chi_{c1}$ channels share the same factorization
framework, the same resonance structure, and largely correlated parametric
dependences. The value $R_{\chi_{c1}/\chi_{c0}}^{\pi\pi}\approx 1.38$
reflects the common resonant production mechanism through the $\rho$ meson,
which largely avoids the state-specific suppression mechanisms
(color-singlet for $\chi_{c0}$, vanishing leading-twist for $\chi_{c1}$)
that operate in the direct two-body transition. A measurement of
$R_{\chi_{c1}/\chi_{c0}}^{\pi\pi}$ by LHCb would directly test this
picture. The predicted large longitudinal polarization fraction
$f_L \approx 93\%$ for the $\chi_{c1}$ channels provides an additional,
independent experimental handle through angular analysis.

\begin{table}[htbp]
\centering
\caption{Comparison of our quasi-two-body results under the narrow-width approximation (NWA) with the two-body
PQCD predictions from Ref.~\cite{Liu:2025}.}
\label{tab:compare}
\small
\renewcommand{\arraystretch}{1.3}
\begin{tabular*}{\textwidth}{@{\extracolsep{\fill}}l c c@{}}
\hline\hline
Decay mode & This work (NWA) & Ref.~\cite{Liu:2025} (two-body) \\
\hline
$\mathcal{B}(B_c^+ \to \chi_{c0} \rho^+)$ & $7.33^{+2.72}_{-2.15} \times 10^{-4}$ & $2.97 \times 10^{-4}$ \\[4pt]
$\mathcal{B}(B_c^+ \to \chi_{c1} \rho^+)$ & $10.15^{+4.50}_{-3.25} \times 10^{-4}$ & $1.40 \times 10^{-3}$ \\[4pt]
$R_{\chi_{c1}/\chi_{c0}}$ & $1.38$ & $4.7$ \\
\hline\hline
\end{tabular*}
\end{table}

Using the LHCb measurement $\mathcal{B}(B_c \to J/\psi \pi^+\pi^0)_{\rho(770)}
= 6.33 \times 10^{-4}$~\cite{LHCb:2023} as a benchmark, we compare the
$\rho(770)$ components, defining
$R_{J/\psi}^{\chi_{c0}} \equiv \mathcal{B}(\chi_{c0}\pi\pi)_{\rho(770)}/
\mathcal{B}(J/\psi\pi\pi)_{\rho(770)}^{\rm exp} \approx 0.92$ and
$R_{J/\psi}^{\chi_{c1}} \approx 1.24$. Both are quasi-two-body decays
through the $\rho(770)$ resonance, directly comparable without the narrow-width
approximation. The PQCD prediction for the $J/\psi$ channel~\cite{Wang:2026}
is $4.44\times10^{-4}$, about $30\%$ below the experimental value; if a
similar pattern holds for $\chi_{cJ}$, the actual yields could be
correspondingly larger than our predictions. Normalized to the same $J/\psi\pi^+\pi^0$ benchmark, the $K\pi$ channels give
$R_{J/\psi}^{\chi_{c0},K\pi} \approx 0.23$ and
$R_{J/\psi}^{\chi_{c1},K\pi} \approx 0.31$.

Under the narrow-width approximation, with $\mathcal{B}(\rho \to \pi\pi)
\approx 100\%$ and $\mathcal{B}(K^* \to K\pi) \approx 100\%$~\cite{PDG:2024}, the two-body
branching fractions are listed in Table~\ref{tab:nwa}:
$\mathcal{B}(B_c^+ \to \chi_{c0} \rho^+) \simeq 7.33^{+2.72}_{-2.15}
\times 10^{-4}$ and $\mathcal{B}(B_c^+ \to \chi_{c1} \rho^+) \simeq
10.15^{+4.50}_{-3.25} \times 10^{-4}$, with the $K^{*+}$ channels at
$1.49^{+0.55}_{-0.44} \times 10^{-4}$ and $1.94^{+0.83}_{-0.62} \times
10^{-4}$.

At present, none of these channels have been measured. The predicted
branching ratios at $10^{-4}$ for the dominant $\rho(770)$ channels
are comparable to the observed $B_c \to J/\psi \pi^+$ and
$B_c \to J/\psi K^+$ decays, and should be accessible at LHCb through
$\chi_{cJ} \to J/\psi \gamma$ with $J/\psi \to \mu^+\mu^-$. The ratios
$R_{\chi_{c1}/\chi_{c0}}$ is particularly suited for
early measurement, since $B_c$ production and $J/\psi$ reconstruction
systematics largely cancel. The large $f_L \approx 93\%$ provides an
additional handle through angular analysis.

\section{Conclusion}\label{sec:conclusion}

We have studied the quasi-two-body decays $B_c^+ \to \chi_{c0,c1} [\rho(K^*) \to]
\pi^+\pi^0(K^0\pi^+)$ in the leading-order PQCD framework. The $\pi\pi$ and
$K\pi$ pairs in the final state are produced primarily through the P-wave
resonances $\rho(770)$, $\rho(1450)$, $\rho(1700)$, and $K^*(892)$, whose
line shapes are parametrized using the Gounaris-Sakurai and relativistic
Breit-Wigner models, respectively. The non-perturbative dynamics associated
with the hadronization of the meson pairs is absorbed into the two-meson
distribution amplitudes, with the time-like form factors $F_{\pi\pi}^{\parallel,
	\perp}(\omega^2)$ and $F_{K\pi}^{\parallel,\perp}(\omega^2)$ incorporating
the resonant contributions. For the $\chi_{cJ}$ mesons, both twist-2 and
twist-3 light-cone distribution amplitudes are included~\cite{Liu:2018,Liu:2025}, with the ratio
$F^{\perp}/F^{\parallel} \approx f_V^T/f_V$ adopted for the perpendicular
form factors.

The CP-averaged branching ratios, polarization fractions, and individual
polarization components have been calculated, and the theoretical uncertainties
from the $B_c$ shape parameter $\beta_{B_c}$, the hard scale $t$, the decay
constants $f_M$, and the Gegenbauer moments $a_\rho/a_{K^*}$ have been
analyzed in detail. The $\rho$ channels, dominated by the $\rho(770)$ resonance, yield branching ratios
at the order of $10^{-4}$--$10^{-3}$, with $\mathcal{B}(B_c^+ \to \chi_{c0} \pi^+\pi^0)
= 7.33 \times 10^{-4}$ and $\mathcal{B}(B_c^+ \to \chi_{c1} \pi^+\pi^0)
= 1.02 \times 10^{-3}$. The $\rho(770)$ resonance accounts for approximately
$79\%$ ($77\%$) of the $\chi_{c0}$ ($\chi_{c1}$) total, while $\rho(1450)$ and $\rho(1700)$ contribute at the
$10^{-5}$--$10^{-6}$ level. The coherent sum of the three $\rho$ resonances
exceeds the single $\rho(770)$ contribution by about $26\%$ ($30\%$) for $\chi_{c0}$ ($\chi_{c1}$), reflecting the
constructive interference inherent in the GS parametrization. The Cabibbo-suppressed $K\pi$ channels yield branching ratios at the order
of $10^{-4}$ for $K^0\pi^+$, with $K^+\pi^0$ receiving half of these values
under isospin symmetry.

Since the considered decays are mediated solely by a single tree-level amplitude carrying the real CKM factor $V_{cb}^*V_{uq}$, the direct $CP$ asymmetries vanish identically, and the $CP$-averaged branching ratios presented above coincide with those of the corresponding charge-conjugate modes. Should a non-zero direct $CP$ asymmetry be observed experimentally in any of these channels, it would constitute a clean signal of new physics beyond the Standard Model.

For the $\chi_{c1}$ channels, the longitudinal polarization fraction is
found to be $f_L \approx 93\%$ for the $\rho(770)$-dominated decays,
comparable to the $f_0 \approx 90\%$ observed in $B_c \to J/\psi \pi\pi$
decays~\cite{Wang:2026}, confirming that longitudinal polarization dominance is a generic
feature of quasi-two-body $B_c$ decays in the PQCD approach. As the $\rho$
resonance mass increases, $f_L$ systematically decreases from $93\%$
($\rho(770)$) to $75\%$ ($\rho(1450)$) and $64\%$ ($\rho(1700)$), while
$f_\parallel$ correspondingly grows, reflecting the enhanced twist-3
contributions at larger dipion invariant mass~\cite{Liu:2018}. The ratio between the
$\chi_{c1}$ and $\chi_{c0}$ total branching fractions is
$R_{\chi_{c1}/\chi_{c0}}^{\pi\pi} \approx 1.38$. This value reflects the
fact that in the quasi-two-body framework both $\chi_{c0}$ and $\chi_{c1}$ are
produced predominantly through the same resonant $\rho$ mechanism, so their
relative yield is set by the competition between the charmonium
decay-constant hierarchy and phase space, in
contrast to the direct two-body transition where the two states are
subject to distinct dynamical suppression mechanisms. An experimental
measurement of this ratio would provide a clean discrimination between
these two production regimes. The corresponding ratio in the Cabibbo-suppressed $K\pi$ channel, $R_{\chi_{c1}/\chi_{c0}}^{K\pi} \approx 1.30$, is close to $R_{\chi_{c1}/\chi_{c0}}^{\pi\pi} \approx 1.38$, providing a cross-check of the framework's internal consistency: the near-equality of the spin ratio in two independent final states confirms that it is governed by the charmonium-side dynamics, with only a mild residual sensitivity to the light-meson distribution amplitudes.

Under the narrow-width approximation, the two-body branching fractions for
$B_c^+ \to \chi_{cJ} V$ have been estimated. These provide a useful
reference for comparison with direct two-body calculations, though the
two frameworks differ in both the treatment of the vector meson (off-shell
resonance versus on-shell final-state particle) and the $k_T$ resummation
precision~\cite{Liu:2025}.

At present, none of the $B_c \to \chi_{cJ} \pi\pi(K\pi)$ decay channels
have been measured experimentally. However, the predicted branching ratios at
the $10^{-4}$ level are comparable to those of the observed $B_c \to
J/\psi \pi^+$ and $B_c \to J/\psi K^+$ decays~\cite{LHCb:2016}, and should become accessible
with the full LHCb dataset through the $\chi_{cJ} \to J/\psi \gamma$ decay
chain~\cite{PDG:2024}. The relative ratio $R_{\chi_{c1}/\chi_{c0}}^{\pi\pi}$ is particularly
well-suited for early measurements, as systematic uncertainties related to the $B_c$ production
cross section largely cancel. The predicted large longitudinal polarization
fraction provides an additional experimental handle through angular analysis
of the $\chi_{c1} \to J/\psi \gamma$ decay. These measurements will help clarify the production mechanism of P-wave charmonia in $B_c$ decays.

We note that the present analysis has been performed at leading order (LO)
in the PQCD expansion, i.e.\ $O(\alpha_s)$ in the hard kernel. The iPQCD
framework~\cite{Liu:2023} incorporates
next-to-leading-logarithm $k_T$ resummation~\cite{Liu:2020} and an
improved treatment of charm-quark effects, and the same framework has been
shown to accommodate next-to-leading-order corrections---including vertex
corrections and quark-loop contributions---which can significantly shift
branching ratios for $B_c \to \chi_{cJ}$ transitions. The two-body iPQCD
results of Ref.~\cite{Liu:2025} are obtained at leading order in
$\alpha_s$. Extending the quasi-two-body analysis to the iPQCD level would
allow a consistent side-by-side comparison with those results and would
help disentangle the genuine dynamical differences between the two
factorization frameworks from artifacts of the LO truncation. We leave this improvement to a future study.

Finally, we have focused on the $\chi_{c0}$ and $\chi_{c1}$ states. The
tensor charmonium $\chi_{c2}$ ($J^{PC}=2^{++}$) has been observed by LHCb
in the two-body mode $B_c^+ \to \chi_{c2}\pi^+$~\cite{LHCb:2024}, and its
quasi-two-body decays $B_c^+ \to \chi_{c2} [\rho(K^*) \to] \pi\pi(K\pi)$
are of equal theoretical interest. The $\chi_{c2}$ case is technically more
involved because of its tensor polarization structure and requires the
corresponding two-meson DAs and $\chi_{c2}$ LCDAs~\cite{Liu:2018},
which are beyond the scope of the present work. We plan to address this
channel in a forthcoming publication.

\begin{acknowledgments}
This work is supported by the National Natural Science Foundation of China under Grant No.11047028.
\end{acknowledgments}

\bibliography{DJnotes}

\end{document}